\documentclass[aip, twocolumn, floatfix,10pt]{revtex4-1}

\usepackage{docs}%
\usepackage{bm}%
\usepackage[colorlinks=true,linkcolor=blue]{hyperref}%
\expandafter\ifx\csname package@font\endcsname\relax\else
 \expandafter\expandafter
 \expandafter\usepackage
 \expandafter\expandafter
 \expandafter{\csname package@font\endcsname}%
\fi
\usepackage{graphicx}   
\usepackage{amsmath}    
\usepackage{amssymb}    
\usepackage{xcolor}      
\usepackage{cancel}
\usepackage[colorlinks=true,linkcolor=blue]{hyperref}%
\usepackage[T1]{fontenc}
\usepackage{ae,aecompl}
\usepackage{caption}

\renewcommand{\vec}[1]{\boldsymbol{#1}}
\newcommand{\hvec}[1]{\hat{\boldsymbol{#1}}}
\newcommand{\pd}[2]{\frac{\partial #1}{\partial #2} }
\newcommand{\HALF}{\frac{1}{2}}

\newcommand{\DS}{\displaystyle}
\newcommand{\tens}[1]{\mathsf{#1}}
\newcommand{\tsigma}{\tilde{\sigma}}
\newcommand{\teta}{\tilde{\eta}}

\newcommand{\typeI}{\emph{type P} }

\newcommand{\typeII}{\emph{type N} }

\newcommand{\Pfive}{ ${\cal P}_5$ }
\newcommand{\Pfour}{ ${\cal P}_4$ }

\begin{document}


\def\apj{\rm ApJ}               
\def\apjs{\rm ApJS}               
\def\mnras{\rm MNRAS}
\def\prd{\rm Physical Review D}
\def\aap{\rm Astronomy \& Astrophysics}


\title{Linear Wave Propagation for Resistive Relativistic Magnetohydrodynamics}%

\author{A. Mignone}%
\affiliation{Dipartimento di Fisica, University di Torino, via Pietro Giuria 1,
             I-10125 Torino, Italy}%

\author{G. Mattia}%
\affiliation{Dipartimento di Fisica, University di Torino, via Pietro Giuria 1,
             I-10125 Torino, Italy}%

\author{G. Bodo}%
\affiliation{INAF, Osservatorio Astrofisico di Torino, Strada Osservatorio 20,
             Pino Torinese 10025, Italy}%

\date{\today}%
\revised{\today}%

\begin{abstract}
We present a linear mode analysis of the relativistic MHD equations in the presence of finite electrical conductivity.
Starting from the fully relativistic covariant formulation, we derive the dispersion relation in the limit of small linear perturbations.
It is found that the system supports ten wave modes which can be easily identified in the limits of small or large conductivities.
In the resistive limit, matter and electromagnetic fields decouple and solution modes approach pairs of light and acoustic waves as well as a number of purely damped (non-propagating) modes.
In the opposite (ideal) limit, the frozen-in condition applies and the modes of propagation coincide with a pair of fast magnetosonic, a pair of slow and Alfv\'en modes, as expected.
In addition, the contact mode is always present and it is unaffected by the conductivity.
For finite values of the conductivity, the dispersion relation gives rise to either pairs of opposite complex conjugate roots or purely imaginary (damped) modes.
In all cases, the system is dissipative and also dispersive as the phase velocity depends nonlineary on the wavenumber.
Occasionally, the group velocity may exceed the speed of light although this does not lead to superluminal signal propagation.
\end{abstract}

\maketitle


\section{Introduction}
\label{sec:introduction}
%
%
%

The dynamics of relativistic plasmas is of great interest both in the laboratory, as in the case of laser produced plasmas, and for high energy astrophysics.
The large scale properties of such plasmas can be described by using the magnetohydrodynamics (MHD) approximation, whose relativistic extension has been developed by  \citet{Lichnerowicz_1967} and  \citet{Anile_2005} paralleling the well studied non-relativistic version.
Relativistic MHD (RMHD henceforth) has been employed over the last decades to describe the dynamics of such systems well in their nonlinear regimes, particularly  through the use of numerical simulations and remarkable progresses have been made in the development of numerical methods for the RMHD equations (see, e.g., \cite{Kom_1999, DelZanna_2003, Balsara_2001, Gammie_2003, Mignone_2006, Giacomazzo_2006}).
Even though the ideal limit, where dissipative effects are neglected, captures effectively the dynamics in most of the situations, there are cases in which resistivity plays a fundamental role, magnetic reconnection is a  notable example.
Therefore in the last years a strong interest has been devoted to the resistive RMHD equations and to finding robust and accurate numerical schemes for their solution \cite{Kom_2007, Palenzuela_2009}.

The ideal RMHD linear wave dynamics is well known: just as in the case of classical MHD, the plasma supports slow, fast magnetosonic and Alfv\'en waves and expressions for the wave speeds have been obtained and used, for example, in numerical schemes for relativistic magnetofluid codes, see \cite{Kom_1999, Balsara_2001, DelZanna_2003, Gammie_2003, vHKM_2008, Mignone_2009} and references therein.
A compendium of the properties of such linear waves can be found, for example, in  \citet{KM_2008}. The properties of linear waves in the resistive case are less well known and the purpose of this paper is to give a comprehensive analysis of such waves.
The results presented in this paper, in addition to being of interest per se, can be particularly relevant for the construction of numerical schemes for the resistive RMHD equations.

The propagation of electromagnetic waves in resistive pair plasmas has been presented by \cite{Koide_2008} using a one-fluid theory derived from the relativistic two-fluid equations.
An approximate dispersion relation for the resistive RMHD equations, that considers only transverse wave propagation (i.e., Alfv\'en waves) in the magnetic field direction, has been derived in appendix of \cite{TI_2011} in the development of a numerical scheme.
More recently, a linear analysis of the resistive RMHD equations has been presented by \cite{DelZanna_etal2016} in the context of tearing mode instability by investigating the stability of an initial force-free current field.
In their study, the authors assume an incompressible plasma and neglect Ampere's law by assuming an electric field which includes the usual convective and diffusive contributions.
In the present work, instead, we present an extensive normal mode analysis of the resistive RMHD equations by retaining the complete form of the equations.
In the presence of resistivity the RMHD equations take the form of hyperbolic equations with relaxation terms \cite{Palenzuela_2009, Liu_1987,  Hittinger_2000, Roe_2001}, this leads to several modifications of the wave properties.
In addition to  introducing wave damping (as one would expect), resistivity leads to other qualitative changes in the wave properties as well.
As in all hyperbolic systems with relaxation, we can distinguish two regimes \citep[see, e.g.][]{Hittinger_2000}: at small wavenumbers resistivity tends to be negligible and the system supports standard RMHD waves, i.e. slow, fast magnetosonic and Alfv\'en; at large wavenumbers, instead, Maxwell equations decouple from the fluid equations and the system supports light and sound waves.
For intermediate wavenumbers, connecting these two regimes, the system becomes dispersive. 

The plan of the paper is the following.
In section \ref{sec:equations}, starting from the full covariant form of the  resistive relativistic MHD equations, we carry out the normal mode analysis in the limit of small perturbations and obtain the characteristic polynomial whose roots give the desired dispersion relation.
In section \ref{sec:general}, we provide asymptotic solutions to the dispersion relations in the resistive and ideal limits.
In section \ref{sec:results}, the solutions of the dispersion relation are analyzed for finite values of the conductivity and for different values of the parameters.
Conclusions are finally drawn in section \ref{sec:summary}.

\section{Equations}
\label{sec:equations}
%
%
%

\subsection{The Resistive Relativistic MHD Equations}
%

Our starting point are the covariant equations of resistive relativistic MHD which follow from the conservation of particle number density and stress-energy tensor coupled to the Maxwell's equations of classical electromagnetism, see \cite{Lichnerowicz_1967, Kom_2007,Palenzuela_2009} and references therein.
Using a system of units where $c=4\pi=1$ we have:
\begin{equation}\label{eq:covariant}
  \left\{\begin{array}{lcl}
  \partial_\alpha (nu^\alpha) &= & 0
  \\ \noalign{\medskip}
  \partial_{\beta}\left(  \tens{T}^{\alpha\beta}\right) &=& 0
  \\ \noalign{\medskip}
    \partial_{\beta}    \tens{F}^{\alpha\beta} & = & -J^{\alpha}
  \\ \noalign{\medskip}
    \partial_{\beta}\; ^{*}\tens{F}^{\alpha\beta} & = & 0
  \end{array}\right.
\end{equation}
where $\tens{F}^{\alpha\beta}$ is the electromagnetic tensor ($\tens{F}^{0i} = -\tens{F}^{i0} = -E_i$, $\tens{F}^{ij} = -\epsilon^{ijk}B_k$), $^{*}\tens{F}^{\alpha\beta}$ is its dual and $J^{\alpha}$ is the four-current vector.

The stress-energy tensor for the composite system fluid+electromagnetic fields can be written as $T^{\alpha\beta} = T_{\rm fluid}^{\alpha\beta} + T_{\rm em}^{\alpha\beta}$ where
\begin{equation}
  \left\{\begin{array}{lcl}
    \tens{T}^{\alpha\beta}_{\rm fluid} & = & w u^{\alpha}u^{\beta} - 
      p\tens{g}^{\alpha\beta}
\\ \noalign{\medskip}
    \tens{T}^{\alpha\beta}_{\rm em} & = & \tens{g}^{\alpha\mu}
      \tens{F}_{\mu\lambda}\tens{F}^{\lambda\beta} + \frac{1}{4}
      \tens{F}_{\mu\nu}\tens{F}^{\mu\nu}\tens{g}^{\alpha\beta} 
  \end{array}\right.
\end{equation}
are, respectively, the fluid and electromagnetic tensors, $w$ is the gas enthalpy, $u^\alpha=\gamma(1,\vec{v})$ is the fluid four-velocity, $p$ is the gas pressure and $g^{\alpha\beta}$ is the metric tensor.

The explicit form of the four-current vector is defined by Ohm's law and accounts only for the plasma resistivity $\eta=1/\sigma$, where $\sigma$ is the electrical conductivity \cite{Lichnerowicz_1967, Kom_2007}:
\begin{equation}
 J^{\alpha} = \frac{1}{\eta}\tens{F}^{\alpha\mu}u_{\mu} + q_{0}u^{\alpha} \,,
\end{equation}
where $q_{0} = -J^\alpha u_\alpha$ is the electric charge density in the fluid rest frame.
Note that the fluid charge $q$ and current density $\vec{J}$ in the lab frame are respectively given by the temporal and spatial components of the four-current:
\begin{align}
  q       &\equiv J^0 = \sigma(\vec{E}\cdot\vec{u}) + q_0\gamma \label{eq:charge}
  \\ \noalign{\medskip}
  \vec{J} &\equiv J^i = \gamma\sigma\left[\vec{E} + \vec{v}\times\vec{B}
                              - (\vec{E}\cdot\vec{v})\vec{v}\right]
                              + q\vec{v} \,.
  \label{eq:J}                              
\end{align}

Projecting Eqs. (\ref{eq:covariant}) in the directions parallel and perpendicular to any time-like vector $n^\mu$, we obtain the three-dimensional form of the resistive relativistic magnetohydrodynamics (RRMHD henceforth) which, after simple manipulations, can be written as
\begin{align}
  \pd{(\rho\gamma)}{t} + \nabla\cdot(\rho\gamma\vec{v}) &= 0
  \label{eq:rrmhd:rho} \\ \noalign{\medskip}
    \pd{}{t}(w\gamma^2\vec{v})
  + \nabla\cdot(w\gamma^2\vec{v}\vec{v})
  + \nabla p
  &= q\vec{E} + \vec{J}\times\vec{B}
  \label{eq:rrmhd:v} \\ \noalign{\medskip}
    \DS \pd{\vec{B}}{t} + \nabla\times\vec{E}  &= 0
  \label{eq:rrmhd:B} \\ \noalign{\medskip}
    \DS \pd{\vec{E}}{t} - \nabla\times\vec{B}  &=  -\vec{J}
  \label{eq:rrmhd:E} \\ \noalign{\medskip}
  \pd{}{t}(w\gamma^2 - p) + \nabla\cdot(w\gamma^2\vec{v}) &= 
     \vec{J}\cdot\vec{E}
  \label{eq:rrmhd:En} 
\end{align}
where $\rho = nm$ is the rest-mass density, $\gamma=(1-\vec{v}^2)^{-\HALF}$ is the fluid Lorentz factor, $\vec{v}$ is the fluid velocity, $\vec{E}$ and $\vec{B}$ are the electric and magnetic field vectors, $w$ and $p$ are the gas enthalpy and pressure, respectively.

The temporal components of the third and fourth equation in (\ref{eq:covariant}) yield the time-independent Maxwell's relations for the field divergences,
\begin{equation}\label{eq:Maxwell_div}
  \nabla\cdot\vec{E} = q\,,\qquad
  \nabla\cdot\vec{B} = 0\,.
\end{equation}
Finally, an equation of state (EoS), in the form $w=w(\rho,p)$, must be provided for appropriate closure.

\subsection{Normal Mode Analysis}
%

The equilibrium state consists of a homogeneous plasma at rest with constant density and pressure $\rho_0$ and $p_0$, respectively.
The system is threaded by a constant and uniform magnetic $\vec{B}_0$ while the electric field must vanish in this frame: $\vec{E}_0 = \vec{0}$.

Equations (\ref{eq:rrmhd:rho})--(\ref{eq:rrmhd:En}) are linearized assuming plane wave perturbations in the form $V_1\propto \epsilon e^{i(\vec{k}\cdot\vec{x} - \omega t)}$, where $V$ is any of the fluid variables, $\epsilon$ is a small amplitude, $\omega$ is the (complex) frequency and $\vec{k}$ is the wavevector.
By retaining only terms of order one, we have
\begin{equation}\label{eq:linearized_RRMHD}
 \left\{
  \begin{array}{lcl}
  -i\omega \rho_1 + i\rho_0 \vec{k}\cdot\vec{v}_1 & = & 0
  \\ \noalign{\medskip}
   -i\omega w_0\vec{v}_1 + i\vec{k}p_1 & = & \vec{J}_1\times\vec{B}_0 
  \\ \noalign{\medskip}
   -i\omega \vec{B}_1 + i\vec{k}\times\vec{E}_1 & = & 0
  \\ \noalign{\medskip}
   -i\omega \vec{E}_1 - i\vec{k}\times\vec{B}_1 & = & -\vec{J}_1
  \\ \noalign{\medskip}
 - i\omega \left[(w'_{p}-1)p_1 + w'_{\rho}\rho_1\right]
                +  w_0i\vec{k}\cdot\vec{v}_1 &=& 0 \,.
 \end{array}\right.
\end{equation}
Here $\vec{J}_1 = \sigma[\vec{E}_1 + \vec{v}_1\times\vec{B}_0]$ is the perturbation of the current density.
From the third equation, we always have $\vec{B}_1\cdot\vec{E}_1=0$ that is, magnetic and electric field perturbations are always orthogonal.
In addition, the divergence-free condition for magnetic field requires $\vec{k}\cdot\vec{B}_1 = 0$.
Also, the Lorentz factor is a second-order quantity ($\gamma\approx O(\epsilon^2)$) and the charge density $q\approx i\vec{k}\cdot\vec{E}_1$ appears only through second (or higher) order terms in $\epsilon$.
Both quantities, therefore, can be neglected.

Without loss of generality, the equilibrium magnetic field is taken to lie in the $x-y$ plane: $\vec{B}_0 = (B_{0x},\, B_{0y},\, 0)$ and we the wavevector $\vec{k}$ along the $x$ direction, $\vec{k}\equiv k \hvec{e}_x$.
The linearized RRMHD equations (\ref{eq:linearized_RRMHD}) can then be written as a homogenous $10\times10$ linear system:
\begin{equation}\label{eq:linear_system}
  \tens{A}\left(\begin{array}{l}
  \rho_1       \\ \noalign{\medskip}
  \vec{v}_1    \\ \noalign{\medskip}
  \vec{B}_{t1} \\ \noalign{\medskip}
  \vec{E}_1    \\ \noalign{\medskip}
  p_1
  \end{array}\right) = 0 \,
\end{equation}
where the matrix $\tens{A}$ is given, in compact form, by 
\begin{equation}\label{eq:Jacobian2}
\tens{A} = \left[ \begin {array}{cccccc}
 -\lambda & \rho_0\vec{e}_x^\top & 0           & 0           & \vec{0}^\top       & 0            \\ \noalign{\medskip}
  \vec{0} & \tens{T}             & \vec{0}     & \vec{0}     & \tens{M}           & \vec{e}_x^\top    \\ \noalign{\medskip}
    0     & \vec{0}^\top         & -\lambda    & 0           & -\hvec{e}_z^\top  & 0           \\ \noalign{\medskip}
    0     & \vec{0}^\top         & 0           & -\lambda    &  \hvec{e}_y^\top  & 0          \\ \noalign{\medskip}
  \vec{0} & -\tens{M}             & -\hvec{e}_z &  \hvec{e}_y & \tens{D}           & \vec{0}    \\ \noalign{\medskip}
  - \lambda w'_{0,\rho} & w_0\vec{e}_x^\top  & 0 & 0 & \vec{0}^\top  & -\lambda (w'_{0,p}-1)
\end {array} \right] 
\end{equation}
In the previous expression $\lambda = \omega/k \in\mathbb{C}$ is the (complex) eigenvalue while $\tens{T}$, $\tens{M}$ and $\tens{D}$ are $3\times3$ matrices with components
\begin{equation}
  \tens{T}_{ij} = -\left(w_0\lambda + i\tsigma B_0^2\right)\delta_{ij} + i\tsigma B_{0i}B_{0j}
\end{equation}
and
\begin{equation}
  \tens{M}_{ij} = i\tsigma\varepsilon_{ijk}B_{0k} \,,\quad
  \tens{D}_{ij} = {\rm diag}(-\lambda + i\tsigma) \,.
\end{equation}
where $\epsilon_{ijk}$ is the Levi-Civita symbol.
Note that the wavenumber and the conductivity always enter through the combination $\tsigma = \sigma/k$.

After straightforward algebra, the characteristic polynomial of (\ref{eq:Jacobian2}) can be written as
\begin{equation}\label{eq:charP}
  {\cal P}(\lambda) = \lambda{\cal P}_5 (\lambda) {\cal P}_4 (\lambda) \,,
\end{equation}
where ${\cal P}_5(\lambda)$ and ${\cal P}_4(\lambda)$ are given by 
\begin{equation}\label{eq:P5}
  \begin{split}
  {\cal P}_5(\lambda) =&\quad   \lambda^5
              + i\tsigma(u_A^{2} + 1)\lambda^4
              - (a^2 + 1)\lambda^3   \\
              & - i\tsigma(a^{2}u_A^{2}\cos^{2}\theta + a^{2} + u_A^{2})\lambda^2 \\
              &  + a^{2}\lambda        
                + i\tsigma a^2 u_A^2\cos^{2}\theta \,,
  \end{split}              
\end{equation}
and
\begin{equation}\label{eq:P4}
  \begin{split}
  {\cal P}_4(\lambda) =&\quad  \lambda^4 + i\tsigma(u_A^2 + 2)\lambda^3
                          - \Big[(u_A^{2} + 1)\sigma^2 + 1\Big]\lambda^2  \\
                       &   - i\tsigma(u_A^2 + 1)\lambda 
                          + \tsigma^{2}u_A^2\cos^{2}\theta \,.
  \end{split}                          
\end{equation}
Note that ${\cal P}_4(\lambda)$ could have been directly obtained from the sub-matrix involving only the equations for $v_{z1},\, B_{z1},\, E_{x1},\, E_{y1}$ which are not coupled to the remaining variables.

Equations (\ref{eq:P5}) and (\ref{eq:P4}) have been expressed in terms of the four parameters $a^2$, $u_A^2$, $\theta$ and $\tsigma$ which we now briefly describe.

\begin{itemize}
\item
The first parameter, $a^2$, defines the square of the sound speed which can be defined in terms of the derivatives of the gas enthalpy $w$:
\begin{equation}\label{eq:a2}
 a^2 = \frac{w_0 - \rho_0 w'_{0,\rho}}{w'_{0,p} - 1}\frac{1}{w_0}
\end{equation}
For an ideal gas, $w_0 = \rho_{0} + \Gamma p_{0}/(\Gamma-1)$ so that the sound speed becomes $a = \sqrt{\Gamma p_0/w_0}$, where $\Gamma$ is the specific heat ratio.
Note that ${\cal P}_4$ is independent of the sound speed.

\item
The second parameter is the magnetization $u_A^2 = B_0^2/w = v_A^2/(1-v_A^2)$ where
\begin{equation}\label{eq:va}
  v_A = \frac{|B_0|}{\sqrt{w_0 + B_0^2}}
\end{equation}
reduces to the Alfv{\'e}n velocity in case of parallel propagation.

\item
The third parameter is the angle $\theta$ between the magnetic field and the wavevector:
\begin{equation}
  \theta = \arctan\left(\frac{B_{0y}}{B_{0x}}\right)\,.
\end{equation}

\item
Finally, the fourth parameter is $\tsigma=\sigma/k$.
\end{itemize}

The zeros of the characteristic polynomial give the desired dispersion relation.
From Eq. (\ref{eq:charP}) we immediately see that ${\cal P}(\lambda)$ possesses one trivial root $\lambda=0$ which corresponds to the contact (or entropy) mode.
The other propagation modes are given by the roots \Pfive and \Pfour.

While some general properties of the solution can be established by inspecting the two polynomials (section \ref{sec:general}), the actual eigenmodes and their dependency on the parameters has to be investigated numerically (section \ref{sec:results}).

\section{General Properties of the Solution}
\label{sec:general}
%
%
%

In general, the eigenvalues $\lambda$ of the system are complex quantities and the real part identifies the phase velocity, i.e., $v_p \equiv \Re(\lambda)$ while the damping rate is proportional to the imaginary part through $-k\Im(\lambda)$.

By taking the complex conjugate of \Pfive or \Pfour, it is easily seen that if $\lambda$ is a solution then the opposite of its complex conjugate, $-\bar{\lambda}$, is also a solution.
Thus roots with non-zero real part must always come as pairs of left- and right-going propagating waves with equal damping rates.
Solution modes of this kind, with non-zero phase velocity, will be labeled \typeI modes.
In addition, as shown in Appendix \ref{app:imag_sol}, \Pfive should always admit a strictly imaginary solution ($\Re(\lambda)=0$) which corresponds to a purely damped, non-propagating mode.
Likewise, \Pfour always has (at least) two imaginary solutions.
Solution modes of this kind will be labeled as \typeII modes.

As we shall see, the system is dissipative since $-\Im(\lambda) > 0$ and also dispersive since the phase velocity depends nonlinearly on $\tsigma=\sigma/k$ and therefore on the wavenumber $k$.
The group velocity can be calculated directly using
\begin{equation}\label{eq:group_vel}
  v_g \equiv \frac{d\omega}{dk}
       =  -\frac{d(\lambda/\tsigma)}{d\tsigma} \tsigma^2 \,.
\end{equation}
Near degenerate points (roots with multiplicity two or higher), Eq. (\ref{eq:group_vel}) can occasionally exceed unity and the system presents peculiarities of anomalous dispersion (regions where the group velocity becomes superluminal).
This, however, does not violates causality as we discuss in Section \ref{sec:signal_velocity}.

In the next sub-sections, we derive analytical expressions which hold in the limit of small $\tsigma$ (the resistive limit) and large $\tsigma$ (ideal limit).
We point out that the resistive limit can be obtained by either fixing the wave number and letting $\sigma\to0$ or, alternatively, by fixing the conductivity and considering large wavenumbers.
Conversely, the ideal limit is recovered for large value of $\sigma$ (at fixed wavelength) or for small wavenumbers (at fixed $\sigma$).

%
%


\subsection{Resistive Limit ($\tsigma\to0$).}
\label{sec:resistive_limit}
%
%
%
In the $\tsigma\to 0$ limit one can easily show that \Pfive simplifies to
\begin{equation}
  {\cal P}_5^{(\tsigma\to0)} =
  \lambda\Big[\lambda^4 - (a^2 + 1)\lambda^2 + a^2\Big] = 0
\end{equation}
whose solutions are
\begin{equation}\label{eq:P5sol_sigma0}
  \lambda_{1} = 0\qquad\lambda_{2,3} = \pm a\qquad\lambda_{4,5} = \pm 1 \,.
\end{equation}
The solutions are thus given by four propagation modes (a pair of acoustic waves and a pair of light modes) and a non-propagating mode.
This is not surprising since, for $\tsigma\to 0$ (infinite resistivity limit), electromagnetic waves and fluid motion are no longer coupled.

Likewise, in the resistive limit, \Pfour reduces to:
\begin{equation}
   {\cal P}_4^{(\tsigma\to0)} = \lambda^{2}(\lambda^{2} - 1) = 0
\end{equation}
with solutions
\begin{equation}\label{eq:P4sol_sigma0}
  \lambda_{6,7} = 0\qquad\lambda_{8,9} = \pm1
\end{equation}
representing a pair of \typeII non-propagating modes and a pair of light waves.

Using a perturbative expansion in $\tsigma$ we find that the first-order correction terms to the eigenvalues are, for the roots of \Pfive:
\begin{equation}\label{eq:P5roots_sigma}
  \begin{array}{lcl}
  \lambda_1 &\approx&\DS
   -i\tsigma u_A^2\cos^2\theta + O(\tsigma^3)
  \\ \noalign{\medskip}
  \lambda_{2,3} &\approx& \DS
  \pm a - i\frac{\tsigma}{2}u_A^2\sin^2\theta  + O(\tsigma^2)
  \\ \noalign{\medskip}
  \lambda_{4,5} &\approx&\DS
  \pm 1 - i\frac{\tsigma}{2} + O(\tsigma^2)
  \end{array}
\end{equation}
valid, of course, only for $\tsigma \ll 1$.
Similarly, we find for \Pfour the regular expansion
\begin{equation}\label{eq:P4roots_sigma}
  \begin{array}{lcl}
  \lambda_{6,7} &=&\DS
  -i\frac{\tsigma}{2}(u_A^2+1)
    \left[1 \pm \sqrt{1-\frac{4u_A^2\cos^2\theta}{(u_A^2+1)^2}}\right]
       + O(\tsigma^3)
  \\ \noalign{\medskip}
  \lambda_{8,9} &=&\DS \pm 1 - i\frac{\tsigma}{2} + O(\tsigma^2) \,.
  \end{array}
\end{equation}
Note that, to first-order in $\tsigma$, the imaginary part of the light modes is $-\tsigma/2$, as also shown by \cite{TI_2011} in the case of parallel propagation.

In our notations, $\lambda_k$ with $k=2,3,4,5,8,9$ are \typeI modes while $\lambda_k$ with $k=1,6,7$ are \typeII modes.
All roots have negative imaginary parts which indicate damping.
The four light modes ($\lambda_{4,5}$ and $\lambda_{8,9}$) behave essentially in the same way and the damping rate varies linearly with the conductivity and it does not depend on the sound speed.
The damping rate of the acoustic wave is proportional to the magnetization and the inclination angle.
The three \typeII modes ($\lambda_1$ and $\lambda_{6,7}$) have different damping rates which all increase with the magnetization ($\propto B_0^2$).
For perpendicular propagation, two of them vanish identically and only one is non-zero.
As we shall see later, this feature holds for any value of $\tsigma$.
Interestingly, it can be shown that the phase velocities of the \typeI modes involve only even powers of $\tsigma$ while the damping term can be expressed as a series of odd powers.

\subsection{Ideal Limit ($\tsigma\to\infty$).}
\label{sec:ideal_limit}
%
%
%

In the limit $\tsigma\to\infty$ we have that \Pfive reduces to the following biquadratic equation:
\begin{equation}\label{eq:P5_infty}
  \begin{split}
  {\cal P}_5^{(\tsigma\to\infty)} =&\quad (u_{A}^{2} + 1)\lambda^4 \\
      & - (a^{2}u_{A}^{2}\cos^{2}\theta + a^{2} + u_{A}^{2})\lambda^2 \\
      & + a^2u_{A}^2\cos^{2}\theta \,.
  \end{split}      
\end{equation}
Eq. (\ref{eq:P5_infty}) admits four propagating modes given by the fast and slow magnetosonic speeds (see, e.g., \cite{DelZanna_2003}):
\begin{equation}\label{eq:fast_and_slow}
  \begin{array}{lcl}
   \lambda_{f\pm} &=&\DS \pm \sqrt{
    \frac{a^2u_A^2\cos^2\theta+a^2+u_A^2+\sqrt\Delta}{2(u_A^2+1)} }
    \\ \noalign{\medskip}
   \lambda_{s\pm} &=&\DS \pm \sqrt{
     \frac{a^2u_A^2\cos^2\theta+a^2+u_A^2-\sqrt\Delta}{2(u_A^2+1)}  } \,,
  \end{array}   
\end{equation}
where $\Delta = (a^2u_A^2\cos^2\theta + a^2 - u_A^2)^2 + 4a^2u_A^2\sin^2\theta$.
Simple differentiation with respect to $\theta$ shows that $\lambda_{f\pm}$ and $\lambda_{s\pm}$ are, respectively, monotonically increasing and decreasing functions of $\theta$ in the range $\theta\in[0,\pi/2]$.
Therefore one always has that $\lambda^2_{s,\pm} \le a^2 \le \lambda^2_{f,\pm}$.
The same condition holds in the non-relativistic limit which is easily obtained by letting $u_A^2 + 1 \to 1$ and $a^2u_A^2 \to 0$.

In the same limit, one finds that \Pfour reduces to the simple quadratic equation
\begin{equation}\label{eq:P4_infty}
 {\cal P}_4^{(\tsigma\to\infty)} =
  \lambda^2 (u_A^2 + 1) - u_A^2\cos^2\theta = 0 \,,
\end{equation}
which admits a pair of Alfv{\'e}n wave solutions 
\begin{equation}\label{eq:Alfven}
   \lambda_{A\pm} = \pm \frac{u_A\cos\theta}{\sqrt{u_A^2 + 1}} \,.
\end{equation}

The asymptotic behavior for large $\tsigma$ can be obtained by conveniently introducing the resistivity parameter $\teta = 1/\tsigma$ and rewriting Eq. (\ref{eq:P5}) and (\ref{eq:P4}) as
\begin{equation}\label{eq:P5_eta}
  \begin{split}
  {\cal P}_5 =& \quad   \teta \lambda^5 + i(u_A^2 + 1)\lambda^4
                  - \teta(a^2 + 1)\lambda^3   \\
              &   - i(a^2u_A^2\cos^2\theta + a^2 + u_A^2)\lambda^2\\
              &   + a^2\lambda\teta + ia^2u_A^2\cos^2\theta  \,,
  \end{split}
\end{equation}
and
\begin{equation}\label{eq:P4_eta}
  \begin{split}
  {\cal P}_4 =& \quad    \teta^2\lambda^4
                 + i\teta(u_A^2 + 2) \lambda^3
                 - (\eta^2 + u_A^2 + 1)\lambda^2 \\
              &  - i\teta(u_A^2 + 1)\lambda
                 + u_A^2\cos^2\theta \,.
  \end{split}                 
\end{equation}

Regular \typeI solutions to these equations, in the limit $\teta\to0$ ($\tsigma\to\infty$), may be found using the same perturbative technique adopted in Section \ref{sec:resistive_limit}.
The result is
\begin{equation}\label{eq:PXroots_eta_regular}
  \begin{array}{lcl}
  \lambda_{f\pm}(\teta) & \approx &\DS \lambda_{f\pm}
 - i\frac{\teta}{2}\frac{(1-\lambda_{f\pm}^2)(\lambda_{f\pm}^2 - a^2)}
                   {\sqrt{\Delta}}
  \\ \noalign{\medskip}
  \lambda_{s\pm}(\teta) & \approx &\DS \lambda_{s\pm}
 - i\frac{\teta}{2}\frac{(1-\lambda_{s\pm}^2)(a^2 - \lambda_{s\pm}^2)}
                   {\sqrt{\Delta}}
  \\ \noalign{\medskip}
  \lambda_{A\pm}(\teta) & \approx & \DS \lambda_{A\pm}
   - i\frac{\teta}{2} \left(1 - \lambda^2_{A\pm}\frac{u_A^2 + 2}{u_A^2 + 1}\right)
  \end{array}                    
\end{equation}
where $\lambda_{f\pm}$ and $\lambda_{s\pm}$ are given by (\ref{eq:fast_and_slow}).
Eq. (\ref{eq:PXroots_eta_singular}) shows that the damping rate of fast and slow modes is proportional to $\teta \equiv k \eta $ and, since $\lambda = \omega / k$, we get that the damping rate is proportional to $\eta k^2$, i.e. it has, as expected, a diffusive behavior.

Equations (\ref{eq:P5_eta}) and (\ref{eq:P4_eta}) also admit asymptotically singular solutions which disappear when $\teta\to0$.
The asymptotic behavior can be recovered by the rescaling method, i.e., by setting $z=\lambda/\teta$ which turns the singular perturbation problem into a regular one.
Solving the regularized problem in $z$ using the perturbative approach and then rewriting the solution in the original variable $\lambda$ yields the three \typeII roots in the \emph{asymptotically singular} (\emph{as}) limit:
\begin{equation}\label{eq:PXroots_eta_singular}
  \begin{array}{lcl}
  \lambda_{as,1} &=&\DS - i\frac{u_A^2 + 1}{\teta}
                   + i\frac{a^2u_A^2\sin^2\theta + 1}{(u_A^2+1)^2}\teta
                   + O(\teta^3)
  \\ \noalign{\medskip}
  \lambda_{as,2} &=&\DS - i\frac{u_A^2 + 1}{\teta}
                   + i\frac{\cos^2\theta}{(u_A^2+1)^2}\teta
                   + O(\teta^3)
  \\ \noalign{\medskip}
  \lambda_{as,3} &=&\DS - \frac{i}{\teta}
                   + i\sin^2\theta \teta
                   + O(\teta^3)
  \end{array}  
\end{equation}
where the first solution ($\lambda_{as,1}$) is the singular root of \Pfive while the remaining two come from \Pfour. 

\subsection{Eigenvectors Structure}
\label{sec:eigenvectors}
%
%

From Eq. (\ref{eq:linear_system}) we can obtain a formal expression for the eigenvectors in terms of the eigenvalue $\lambda$.
A generic eigenvector component represents a perturbation that can be written as $V_1 = |V_A|e^{i(kx - \omega_{R}t + \varphi)} e^{\omega_It}$ where $V_A\in\mathbb{R}$ is the wave amplitude and $\varphi$ is the wave phase.

Whenever a plane wave carries a non-zero density perturbation (compressible mode), we assume $\rho_1=\epsilon e^{i(kx-\omega t)}$ and, after some algebra, we obtain
\begin{equation}\label{eq:eigenvectors_comp}
\left(\begin{array}{l}
  \rho_1       \\ \noalign{\medskip}
  v_{1x}       \\ \noalign{\medskip}
  v_{1y}       \\ \noalign{\medskip}
  v_{1z}       \\ \noalign{\medskip}
  B_{1y}       \\ \noalign{\medskip}
  B_{1z}       \\ \noalign{\medskip}
  E_{1x}       \\ \noalign{\medskip}
  E_{1y}       \\ \noalign{\medskip}
  E_{1z}       \\ \noalign{\medskip}
  p_1
  \end{array}\right)
   =  \rho_1
\left(\begin{array}{c}
  1
  \\ \noalign{\medskip}
  \DS \frac{\lambda}{\rho_0}
  \\ \noalign{\medskip}
  \DS \frac{\lambda \sin\theta\cos\theta\Delta_2}{\rho_0\Delta}
   \\ \noalign{\medskip}
   0
   \\ \noalign{\medskip}
  \DS -\frac{\lambda^2w_0B_0\tsigma\sin\theta}{\rho_0\Delta}
   \\ \noalign{\medskip}
   0
   \\ \noalign{\medskip}
   0
   \\ \noalign{\medskip}
   0
   \\ \noalign{\medskip}
   \DS \frac{\lambda^3w_0B_0\tsigma\sin\theta}{\rho_0\Delta}
   \\ \noalign{\medskip}
   \DS \frac{\lambda^2w_0(\Delta_1 + \Delta_2)}{\rho_0\Delta} 
  \end{array}\right) \,,
\end{equation}
where $\Delta = \Delta_1 + \cos^2\theta\Delta_2$, $\Delta_1 = \lambda w_0(i\lambda^2-\lambda\tsigma - i)$, $\Delta_2 = \tsigma B_0^2(1-\lambda^2)$.
Compressible modes are possible only if $\lambda$ is a root of \Pfive (roots of \Pfour do not involve density perturbations as explained after Eq. \ref{eq:P4}).
From the previous expression it is seen that velocity and magnetic field perturbations lie in the plane defined by $\vec{k}$ and $\vec{B}_0$ whereas the electric field is orthogonal to this plane.
In the infinite conductivity limit, perturbations are real quantities and the resulting expressions are well-behaved yielding the eigenvectors for the fast and slow magnetosonic waves (see Appendix \ref{app:eigenvectors_limits}).
For finite values of $\tsigma$, perturbations become complex quantities and a phase shift appear.
Of particular interest is the case of a purely imaginary eigenvalue, i.e., $\lambda=iY$: Eq. (\ref{eq:eigenvectors_comp}) shows that velocity and electric field perturbations become out of phase by $\pi/2$ with respect to those of density, magnetic field and pressure.

By setting $\rho_1=0$ in Eq. (\ref{eq:linear_system}), only the $4\times4$ sub-system formed by the equations of $\{v_{1z}, B_{1z}, E_{1x}, E_{1y}\}$ has non-trivial solution. 
The incompressible perturbations modes are thus associated with the roots of \Pfour and can be written as
\begin{equation}\label{eq:eigenvectors_incomp}
\left(\begin{array}{l}
  \rho_1       \\ \noalign{\medskip}
  v_{1x}       \\ \noalign{\medskip}
  v_{1y}       \\ \noalign{\medskip}
  v_{1z}       \\ \noalign{\medskip}
  B_{1y}       \\ \noalign{\medskip}
  B_{1z}       \\ \noalign{\medskip}
  E_{1x}       \\ \noalign{\medskip}
  E_{1y}       \\ \noalign{\medskip}
  E_{1z}       \\ \noalign{\medskip}
  p_1
  \end{array}\right)
   =  B_{1z}
\left(\begin{array}{c}
  0
  \\ \noalign{\medskip}
  0
  \\ \noalign{\medskip}
  0
   \\ \noalign{\medskip}
   \DS \frac{1-\lambda^2-i\lambda\tsigma}{i\tsigma B_0\cos\theta}
   \\ \noalign{\medskip}
  0
   \\ \noalign{\medskip}
  1
   \\ \noalign{\medskip}
  \DS \tan\theta\frac{1-\lambda^2-i\lambda\tsigma}{\lambda + i\tsigma}
   \\ \noalign{\medskip}
  \lambda
   \\ \noalign{\medskip}
   0
   \\ \noalign{\medskip}
   0
  \end{array}\right) \,.
\end{equation}
Modes described by Eq. (\ref{eq:eigenvectors_incomp}) propagate fluctuations of velocity and magnetic field components perpendicular to the plane defined by $\vec{k}$ and $\vec{B}_0$.

Limit expressions in the resistive and ideal regimes are reported in Appendix \ref{app:eigenvectors_limits}.

\section{Results}
\label{sec:results}
%
%
%

\begin{figure*}[!pt]
  \centering
  \includegraphics[width=0.9\textwidth]{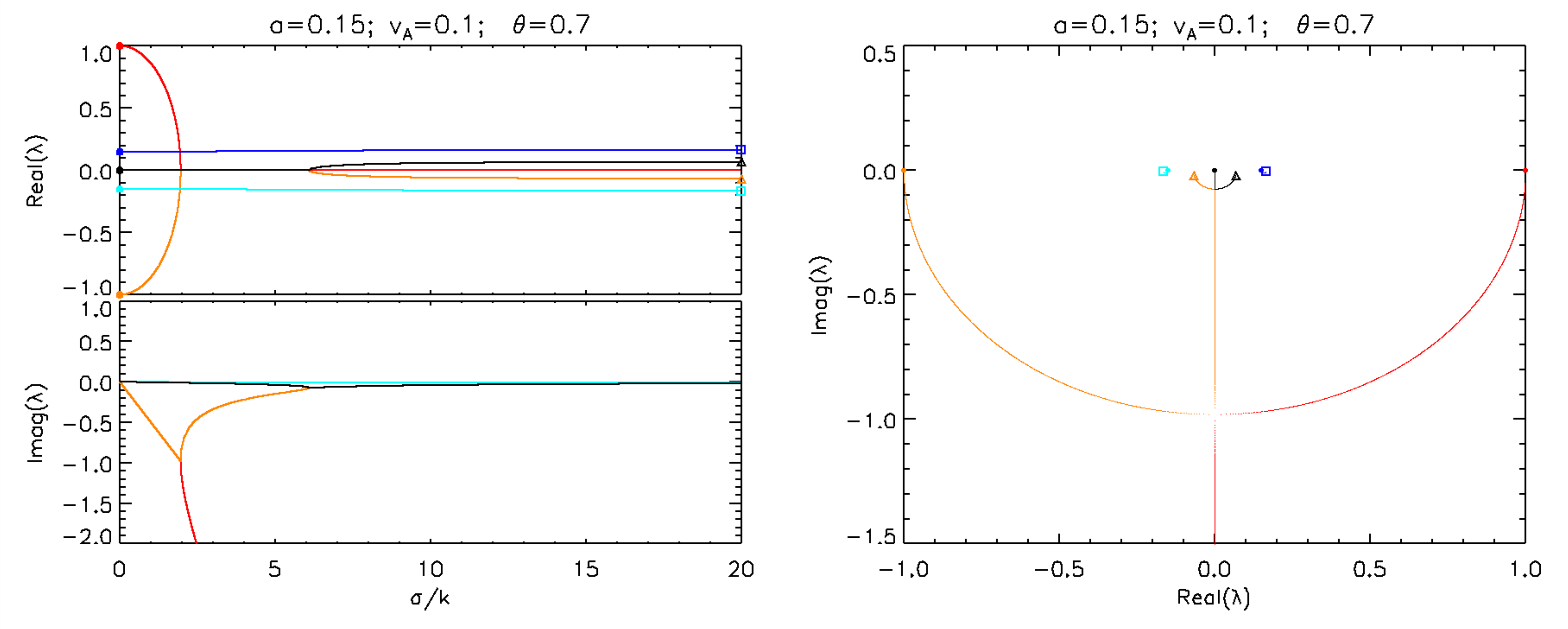}%
  \caption{Roots of ${\cal P}_5$ in the cold gas case ($a=0.15$), low
           magnetization regime ($v_A = 0.1$) and $\theta = 0.7$.
           In the left panel we plot the real and imaginary parts of the 
           solution as functions of $\tsigma\equiv \sigma/k$.      
           The right panels shows the corresponding eigenmodes positions
           in the complex $\lambda$ plane (initial values at $\tsigma = 0$
           are denoted with small filled circles).
           At small values of $\tsigma$, blue and cyan curves denotes the acoustic
           modes while red and orange curves represent the light modes;
           the black line is a purely damped mode.
           At large values of $\tsigma$,  blue and cyan curves tend to the fast
           magnetosonic waves (small squares), black and orange 
           curves approach the slow magnetosonic waves (small triangles) while
           the red line show the rapidly damped mode.}
  \label{img:cold_lowM}
\end{figure*}
We now study in detail the solutions of the characteristic polynomial by exploring the parameter space defined by $a$, $v_A$, $\theta$ and $\tsigma$.

Since neither ${\cal P}_{5}$ nor ${\cal P}_{4}$ have simple analytical solutions for finite value of the conductivity $\tsigma \equiv \sigma/k$, we adopt a numerical approach based on the Durand-Kerner method \cite{DK71} which is widely used for calculating both the real and the complex roots of a univariate polynomial at the same time.
Given a polynomial of $m-$th degree, the Durand-Kerner algorithm iterates on all of the roots $\lambda_i$ (with $i=1,...,m$) simultaneously:
\begin{equation}\label{eq:Durand_Kerner}
 \lambda_i^{(k+1)} = \lambda_i^{(k)} - \DS\frac{{\cal P}_m(\lambda_i^{(k)})}
                             {\DS\prod_{j\neq i}(\lambda_i^{(k)} - \lambda_j^{(*)})} \,,
\end{equation}
where $k$ is the iteration cycle, $\lambda_j^*$ is the most recent updated value ($\lambda^{(*)} = \lambda_j^{(k)}$ if $j>i$ or $\lambda^{(*)}=\lambda_j^{(k+1)}$ otherwise).
The iteration process converges quadratically provided sufficiently close guesses are provided.

Equation (\ref{eq:Durand_Kerner}) is typically solved by fixing $a$, $u_A$ and $\theta$ for different values of the conductivity $\tsigma$.
We start at $\tsigma=0$ where we have exact expressions for the eigenvalues given by Eq. (\ref{eq:P5sol_sigma0}) and (\ref{eq:P4sol_sigma0}), respectively.
These values are then used as guesses to start the iteration cycle for the next value of $\tsigma$.


We first discuss, in sections \ref{sec:resultsP5} and \ref{sec:resultsP4}, the characteristic modes of ${\cal P}_5$ and ${\cal P}_4$ for fixed orientation angle $\theta=0.7 \approx 40^{\circ}$).
Next, in section \ref{sec:resultsTheta}, we examine the behavior of the system at arbitrary angles $\theta$.

As already stated in section \ref{sec:general} we conveniently label \typeI mode pairs of propagating waves with non-zero phase velocity, that is, $\lambda^{(P)} = \pm\Re(\lambda) + i\Im(\lambda)$.
On the contrary, \typeII modes are purely imaginary, non propagating damped modes and have the form $\lambda^{(N)} = i\Im(\lambda)$.
A transition from a \typeI mode to a \typeII mode (e.g. light to purely damped waves) can occur through a \emph{degeneracy point} characterized by a root of multiplicity two.
In these cases, degeneracy points are (by convention) named after the limiting value of the \typeI mode at $\tsigma\to 0$ (for a $P-N$ transition) or $\tsigma\to\infty$ (for a $N-P$ transition).
Likewise, a pair of \emph{degeneracy points} appears in correspondence of two double roots and marks a transition between pairs of \typeI modes (e.g. light-acoustic).

\subsection{Mode Analysis for ${\cal P}_5$}
\label{sec:resultsP5}
%
%

\subsubsection{Results for a Cold Gas.}

We first consider the cold gas case with $a = 0.15$ and study the behavior of the system for different values of the magnetization.


\vspace{10pt}
\paragraph{Low Magnetization ($0.1 \lesssim v_A \lesssim 0.2$).}

In Fig. \ref{img:cold_lowM} we plot the roots of ${\cal P}_5$ for $v_A = 0.1$ and $\theta=0.7$.
In the left panel the real and imaginary parts are plotted as functions of $\tsigma=\sigma/k$ while the right panel gives the path followed in the complex $\lambda$ plane.
The different curves show the five modes which can be easily identified in the limit of zero conductivity (see Eq. \ref{eq:P5sol_sigma0}).
Starting at $\tsigma = 0$, in fact, we have a pair of light modes $\lambda_{2,3}=\pm1$ (red and orange curves in the figure), a pair of acoustic modes $\lambda_{4,5} = \pm a$ (blue and cyan) and a null-mode $\lambda_1 = 0$ (black).
In the limit of small $\tsigma$ our results agree with the expansion given in Eq. (\ref{eq:P5roots_sigma}).

For $0\lesssim \tsigma \lesssim 1.96$, the phase velocities of the light modes decrease (in absolute value) until they become degenerate reaching zero phase speed.
The \emph{light degeneracy point} sets the transition to a pair of \typeII modes and the corresponding formation of a pair of damped standing waves for $1.96 \lesssim \tsigma \lesssim 6$ (red and orange curves on the imaginary axis in the left panel of Fig. \ref{img:cold_lowM}).
As noticed in Section \ref{sec:eigenvectors}, modes with purely imaginary part are characterized by a $\pi/2$ phase shift between velocity and magnetic field perturbations.
The damping rates of the \typeII modes have opposite trend: while one the two modes becomes rapidly suppressed (red), the other one (orange) features a decreasing damping rate until it merges with the purely damped mode (black) at $\tsigma\approx 6$,
This settles the \emph{slow degeneracy point} and the transition to \typeI modes which asymptotically approach a pair of left- and right-going slow magnetosonic waves.

The acoustic modes (blue and cyan), on the other hand, remain always distinct and are characterized by very small damping rates which vanishes as they approach the fast magnetosonic speed in the ideal limit, see Eq. (\ref{eq:P5roots_sigma}).
They also weakly depends  on $\tsigma$.

\paragraph{Moderate Magnetization ($v_A\approx 0.19$)}

\begin{figure*}[!bp]
  \centering
  \includegraphics[width=0.99\textwidth]{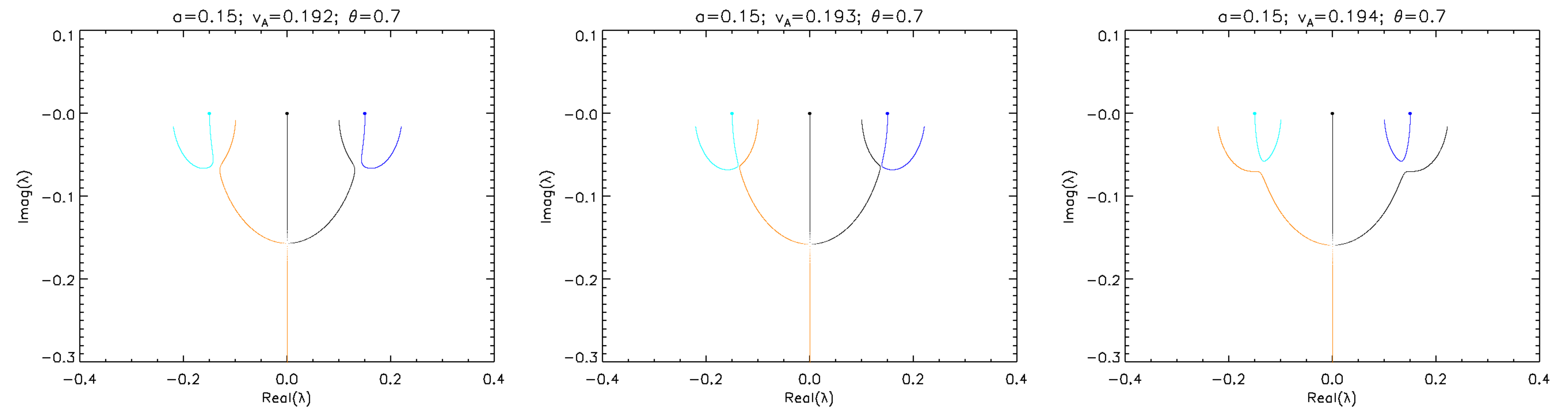}%
  \caption{Merging and asymptotic switch of the light and acoustic modes
           in the complex plane.
           From left to right, the three panels trace the eigenmode position
           in the complex plane for $v_A = 0.191, 0.192$ and $0.193$.
           The double degeneracy point takes place in the middle panel.}
  \label{img:cold_transition1}
\end{figure*}
By increasing the magnetization, the light and acoustic modes move closer in the complex plane.
At $v_A \approx 0.193$ two double roots appear (the \emph{light-acoustic degeneracy point}) and hence the two mode pairs switch their asymptotic branches: the acoustic modes now tend to the slow magnetosonic waves (rather than the fast) while the light modes approach the fast (rather than the slow) modes.
This pattern is best illustrated in Fig. \ref{img:cold_transition1} where the roots are plotted in the complex $\lambda$ plane immediately prior and after the degeneracy, which takes place for $\tsigma\approx 3.8$

\paragraph{High Magnetization ($0.25\lesssim v_A \lesssim 0.41$).}

\begin{figure*}[!pt]
  \centering
  \includegraphics[width=0.9\textwidth]{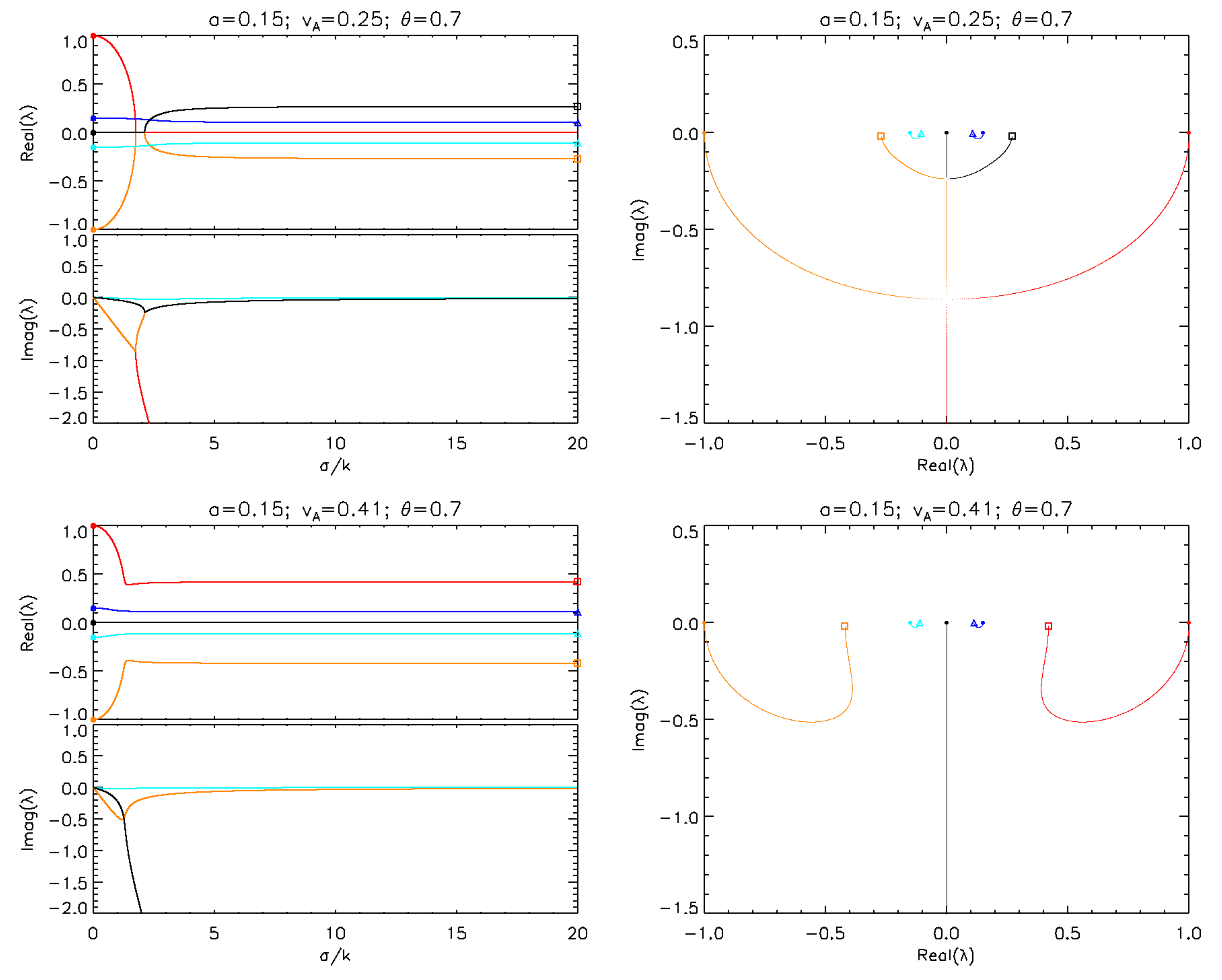}
  \caption{Roots of ${\cal P}_5$ in the cold gas case ($a=0.15$) for
           larger magnetizations corresponding to $v_A= 0.25$ (top panel) and
           $v_A=0.41$ (bottom panel). Plot symbols have the same meaning as
           in Fig. \ref{img:cold_lowM}.}
  \label{img:cold_highM}
\end{figure*}

For $v_A = 0.25$ (top panels in Fig. \ref{img:cold_highM}), the light degeneracy point shifts at slightly smaller value of $\tsigma\approx 1.75$.
Damped standing waves (corresponding to a pair of \typeII modes) form in a much narrower range on the negative imaginary axis.
At $\tsigma \approx 2.1$ we have again a \typeII-\typeI transition through the fast degeneracy point leading to a pair of forward/reverse waves approaching the fast magnetosonic speed (rather than the slow) in the $\tsigma\to\infty$ limit.

When the magnetization is further increased to $v_A=0.41$ (bottom panels) degeneracies are removed and all roots remain distinct for any value of $\tsigma$.
This is best seen in the bottom right panel of Fig. \ref{img:cold_highM} where four \typeI modes (orange, cyan, blue and red) and an isolated \typeII solution are visible.
While the acoustic modes smoothly connect with the slow mode in the ideal limit, the phase velocities of the light waves decrease, in absolute value, to a minimum (found at $\sigma\approx 1.43$) and shortly after rapidly approach the fast magnetosonic speeds.
Finally, the \typeII  mode increases linearly for $\tsigma \lesssim 1$ (see the first equation in \ref{eq:P5roots_sigma}) and then much faster for $\tsigma \gtrsim 1$.

\subsubsection{Results for a Hot Gas}

\begin{figure*}[!pt]
  \centering
  \includegraphics[width=0.9\textwidth]{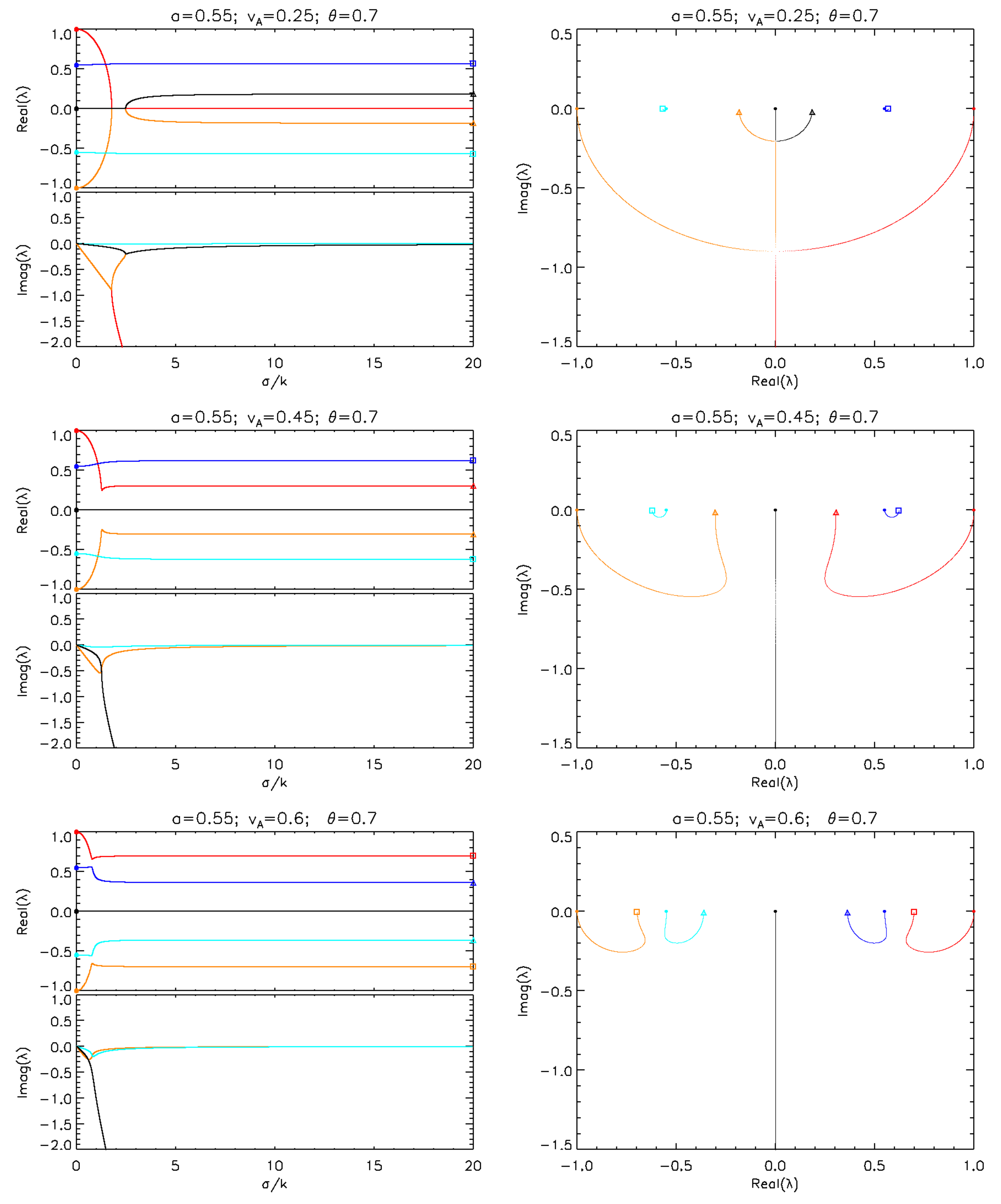}
  \caption{\footnotesize Roots of ${\cal P}_5$ in the hot gas case ($a=0.55$)
  and different magnetizations, as reported in the title.
  Plot symbols have the same meaning as in Fig. \ref{img:cold_lowM}.}
  \label{img:hot}
\end{figure*}

Next we increase the sound speed to $a=0.55$ (slightly below the asymptotic value $1/\sqrt{3}$), in order to investigate relativistic thermodynamics effects.
Eigenvalues are plotted in the six panels of Fig. \ref{img:hot} for increasing values of the magnetization (from top to bottom, $v_A=0.25,\, 0.45$ and $0.6$, respectively).
Although the qualitative behavior is essentially the same one identified for the cold gas case, few differences are discernible.

For $v_A \lesssim 0.25$ (top panels) we again have, for increasing $\tsigma$, two light waves followed by a pair of \typeII modes and then a pair of slow magnetosonic waves.
The damped standing waves are delimited by the two degeneracy points around $\tsigma\approx 1.77$ and $\tsigma\approx 2.48$.
Acoustic modes (blue and cyan) show a weak dependence of the conductivity and smoothly connect to the fast magnetosonic waves.

At $v_A=0.45$ (middle panels), degeneracies have been removed and we have again five distinct modes (4 \typeI solutions and 1 \typeII mode).
Light and slow magnetosonic waves are connected continuously and so are the acoustic-fast magnetosonic waves.
The non-propagating \typeII mode (black) becomes quickly damped as $\tsigma$ increases.

Finally, when the magnetization reaches $v_A=0.6$ (bottom panels), light and acoustic modes swap their asymptotic behavior through a double degeneracy point: the light (acoustic) modes approach the fast (slow) magnetosonic speeds.
The \typeII mode shows the same features as in the cold gas case as its asymptotic behavior (see $\lambda_{as,1}$ in Eq. \ref{eq:PXroots_eta_singular}) is independent of the sound speed.


\subsection{Mode Analysis for ${\cal P}_4$}
\label{sec:resultsP4}
%

\begin{figure*}[!pt]
  \centering
  \includegraphics[width=0.9\textwidth]{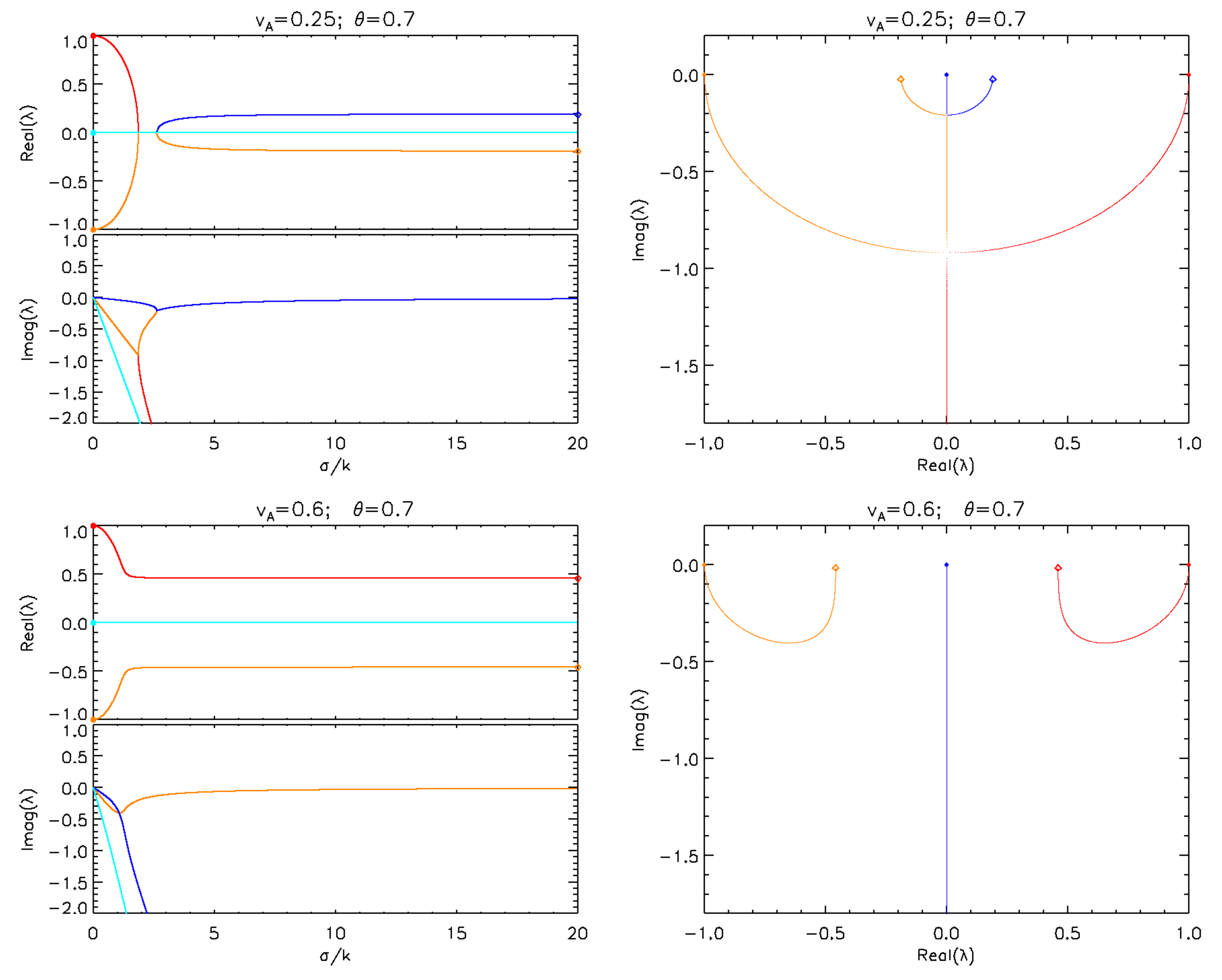}%
  \caption{Roots of ${\cal P}_4$ for $\theta=0.7$ and $v_A = 0.25$ (top panels) or
           $v_A=0.6$ (bottom panels).
           The rapidly damped mode (cyan) has been omitted from the right
           panel for the sake of clarity.}
  \label{img:P4}
\end{figure*}
Since ${\cal P}_4$ does not depend on the sound speed, it suffices to consider different values of $v_A$. 
The overall behavior of roots is qualitatively similar albeit simpler than the cases discussed above.
This is shown in the two panels of Fig. \ref{img:P4} for $v_A=0.25$ (top) and $v_A = 0.6$ (bottom).
For small values of $\tsigma$, we always have two damped light modes (red and orange curves) and a pair of purely damped \typeII modes with different imaginary parts (blue and cyan lines).
To first-order in $\tsigma$, these modes are given by the regular expansions in Eqs. (\ref{eq:P4roots_sigma}).
The mode with larger damping (cyan) remains always distinct and it coincides with $\lambda_7$ in Eq. (\ref{eq:P4roots_sigma}) or $\lambda_{as,3}$ in Eq. (\ref{eq:PXroots_eta_singular}) in the small or large $\tsigma$ limits, respectively.

For $v_A=0.25$, the phase velocity of the light modes decreases (in absolute value) and a \typeI-\typeII transition takes place at the light degeneracy point around $\sigma\approx 1.86$ (top panels).
Here the imaginary part of the light modes is intermediate between the two damped modes, i.e., $\Im(\lambda_6)<\Im(\lambda_{8,9}) < \Im(\lambda_7)$.
A pair of damped standing waves forms for a narrow value range of $\tsigma$ ($1.86 \lesssim \tsigma\lesssim 2.62$) and while one of the two modes becomes rapidly suppressed, the other one (orange) features a smaller damping rate.
At $\tsigma\approx2.62$ we have a second degeneracy (the Alfv\'en degeneracy point) accompanied by a \typeII-\typeI mode transition.
Increasing $\tsigma$ leads to the appearance of Alfv\'en waves.

For $v_A=0.6$, both degeneracies have been removed and all roots are now distinct: a pair of smoothly connected light-Alfv\'en modes and a pair of damped modes with rapidly growing damping rates (bottom panels in Fig. \ref{img:P4}).
The two light modes decrease their speed of propagation until a minimum value in the range $1<\sigma<2$, and then approach the Alfv\'en velocity as $\sigma\to\infty$.
In the same limit, the asymptotic expression for the \typeII modes is given by the singular perturbation solution given in Eq. (\ref{eq:PXroots_eta_singular}).

\subsection{Dependency on the Angle $\theta$}
\label{sec:resultsTheta}
%

While in the previous sections the angle between the wavevector $\vec{k}$ and the magnetic field $\vec{B}$ has been fixed to $\theta=0.7$, we now explore the effect of different orientation angles.
We first consider, in the next two subsections, the limiting cases corresponding to parallel and perpendicular propagation and leave the discussion at arbitrary angles to the last subsection.

\subsubsection{Parallel Propagation ($\theta = 0$).}

\begin{figure*}[!pt]
  \centering
  \includegraphics[width=0.99\textwidth]{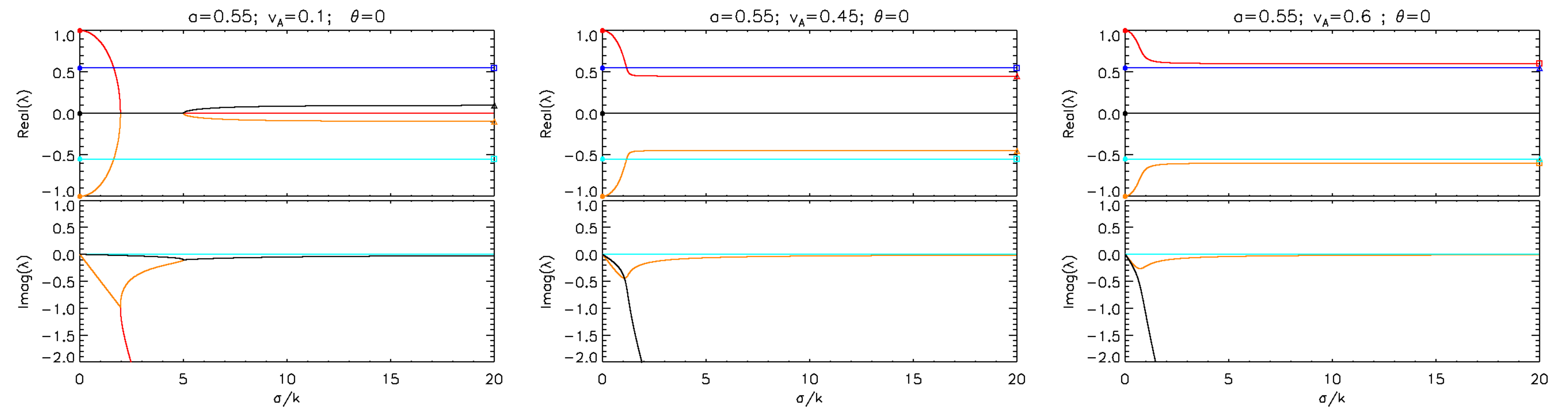}
  \caption{Eigenmodes of ${\cal P}_5$ and ${\cal P}_4$ in the case of parallel
   propagation ($\theta=0$) as a function of $\tsigma$ .
   Note that while the acoustic modes (blue and cyan curves) are roots of
   ${\cal P}_5$ only, the other modes (red, orange and black) are common roots
   to \emph{both} ${\cal P}_5$ and ${\cal P}_4$. 
   The non-propagating and rapidly damped mode of ${\cal P}_4$ has been
   omitted for clarity.}
  \label{img:theta0}
\end{figure*}

When $\vec{B}$ and $\vec{k}$ are aligned, the two characteristic polynomials simplify to 
\begin{align}
  {\cal P}^{\parallel}_5 &= (\lambda^2 - a^2)
            \Big[\lambda^3 + i\tsigma(1 + u_A^2)\lambda^2
                 -\lambda - i\tsigma u_A^2\Big]
  \label{eq:P5_theta0}                  
  \\ \noalign{\medskip}
  {\cal P}^{\parallel}_4 & = (\lambda + i\tsigma)
             \Big[\lambda^3 + i\tsigma(1 + u_A^2)\lambda^2
                 -\lambda - i\tsigma u_A^2\Big]
  \label{eq:P4_theta0}
\end{align}
Eq. (\ref{eq:P5_theta0}) always admits the solutions $\lambda = \pm a$ which show that acoustic wave propagation is unaffected by electrical resistivity. 
Eq. (\ref{eq:P4_theta0}) has the solution $\lambda = -i\tsigma$ which corresponds to the rapidly damped mode (again $\lambda_7$ or $\lambda_{s,3}$ in the opposite limits).
The remaining solutions are given by the roots of the cubic in square bracket which is common to both ${\cal P}_5$ and ${\cal P}_4$ and depend solely on $u_A$.
They reduce to a null mode and a pair of light modes $\lambda={\pm 1,0}$ (for $\tsigma\to0$) or a pair of Alfv\'en waves $\lambda=\pm v_A$ (for $\tsigma\to\infty$).
This result has also been found in the appendix of \cite{TI_2011}.

From the discriminant of the cubic, it is easily found that a pair of \typeII waves joining the light and Alfv\'en degeneracy points (given the black line segment with vanishing real part in the left panel of Fig. \ref{img:theta0}) is found between the two values of $\tsigma$ satisfying
\begin{equation}\label{eq:theta0_double}
  \tsigma^2_c = \frac{-8u_A^4 + 20u_A^2 + 1 \pm (1-8u_A^2)^{3/2}}
                     { 8u_A^2(u_A^2 + 1)^3} \,.
\end{equation}
When $u_A = 1/\sqrt{8}$ ($v_A=1/3$) a triple root $\lambda = -i/\sqrt{3}$ forms at $\tsigma_c \equiv 8\sqrt{3}/9\approx 1.54$.
The degeneracy is then removed when $v_A \ge 1/3$ so that five distinct roots appear with the two light modes always approaching the Alfv\'en velocity while the non propagating mode becoming rapidly damped.
This behavior, shown in the middle and right panels of Fig. \ref{img:theta0}, is also found in classical MHD.

\subsubsection{Perpendicular Propagation ($\theta = \pi/2$).}
\label{sec:perp_prog}

When $\vec{k}$ and $\vec{B}$ are perpendicular, ${\cal P}_5$ reduces to the following expression:
\begin{equation}\label{eq:P5perp}
{\cal P}^{\perp}_5 = \lambda\Big[\lambda^4 + i\tsigma(u_{A}^{2} + 1)\lambda^3
                         - (a^2 + 1)\lambda^2
                         - i\tsigma(a^{2} + u_A^{2})\lambda
                         + a^{2} \Big]
\end{equation}
which always has a vanishing root.
At $\tsigma=0$ we recover the usual pairs of light and acoustic modes while, in the limit $\tsigma\to\infty$, the polynomial inside the square bracket admits the magnetoacoustic wave solution:
\begin{equation}\label{eq:magnetoacoustic}
 \lambda_{f\pm} = \pm\sqrt{\frac{a^2+u_A^2}{u_A^2+1}}
\end{equation}
and a second $\lambda = 0$ solution.
The two vanishing roots at $\tsigma=\infty$ show that the slow magnetosonic modes disappear, as in classical MHD.

\begin{figure}[!ht]
  \centering
  \includegraphics[width=0.45\textwidth]{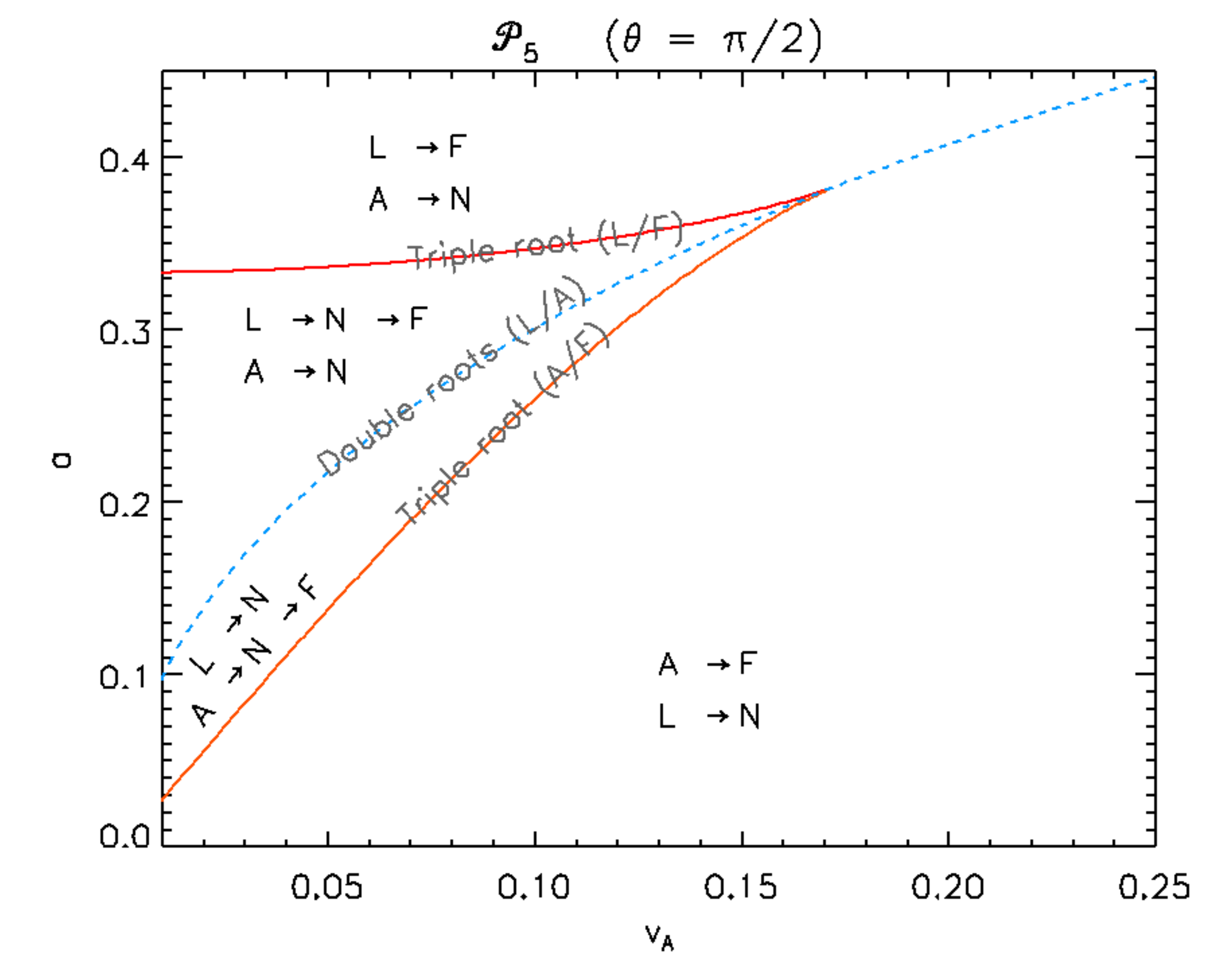}
  \caption{Eigenmodes degeneracies for ${\cal P}_5$ when $\theta = \pi/2$.
  Red and orange curves give the locus of $(v_A, a)$ points where a triple root
  exist (plus and minus sign in Eq. \ref{eq:P5perp_triple_vA}).
  Light (acoustic) waves are never degenerate above (below) the red (orange) curve
  and they smoothly connect to the fast modes in the ideal limit.
  In-between the triple point curves, light (acoustic) waves become degenerate
  for a finite value range of $\tsigma$ if they lie above (below) the blue
  line (Eq. \ref{eq:P5perp_double_root}) but retain the same asymptotic limit.}
  \label{img:P5perp_triple}
\end{figure}
It is possible to show (see Appendix \ref{app:P5perp_triple}) that the quartic inside the square brackets in Eq. (\ref{eq:P5perp}) admits a triple root when
\begin{equation}\label{eq:P5perp_triple_vA}
  v_{A,\pm} = \frac{\sqrt{2}}{4}\sqrt{\frac{B \pm (1-a^2)C^{3/2}}{(a^2+1)^3}}
  \;\quad\mathrm{for}\quad  a < 3-\sqrt{8}\,,
\end{equation}
in correspondence of $\tsigma$ given by Eq. (\ref{eq:P5perp_triple_sigma}).
The coefficients $B$ and $C$ are given immediately after Eqs. (\ref{eq:uA_triple}).
In the $(a,\,v_A)$ plane (see Fig. \ref{img:P5perp_triple}), the two solutions given by Eq. (\ref{eq:P5perp_triple_vA}) define the lower boundary curve above which light modes are no longer degenerate (for $v_A > v_{A,+}$) or the curve below which acoustic modes never degenerate (for $v_A < v_{A,-}$).

Also, a couple of double roots with non-zero phase speed appears when
\begin{equation}\label{eq:P5perp_double_condition}
  v_A = \sqrt{\frac{a}{1+a}}\;\;\;({\rm or}\quad u_A=\sqrt{a})\,,
  \; \quad\mathrm{for}\quad  0 \le a \le 1\,,
\end{equation}
in correspondence of $\tsigma = 2(1-a)/(1+a)$ where
\begin{equation}\label{eq:P5perp_double_root}
  \lambda^\perp_{\pm} = \frac{1}{2}\Big[\pm\sqrt{-a^2 + 6a - 1} - i(1-a)\Big] \,.
\end{equation}
This pair of roots with multiplicity 2 marks a light-acoustic degeneracy point with a corresponding asymptotic switch (similar to the situation illustrated in Fig. \ref{img:cold_transition1}).

\begin{figure*}[!pt]
  \centering
  \includegraphics[width=0.9\textwidth]{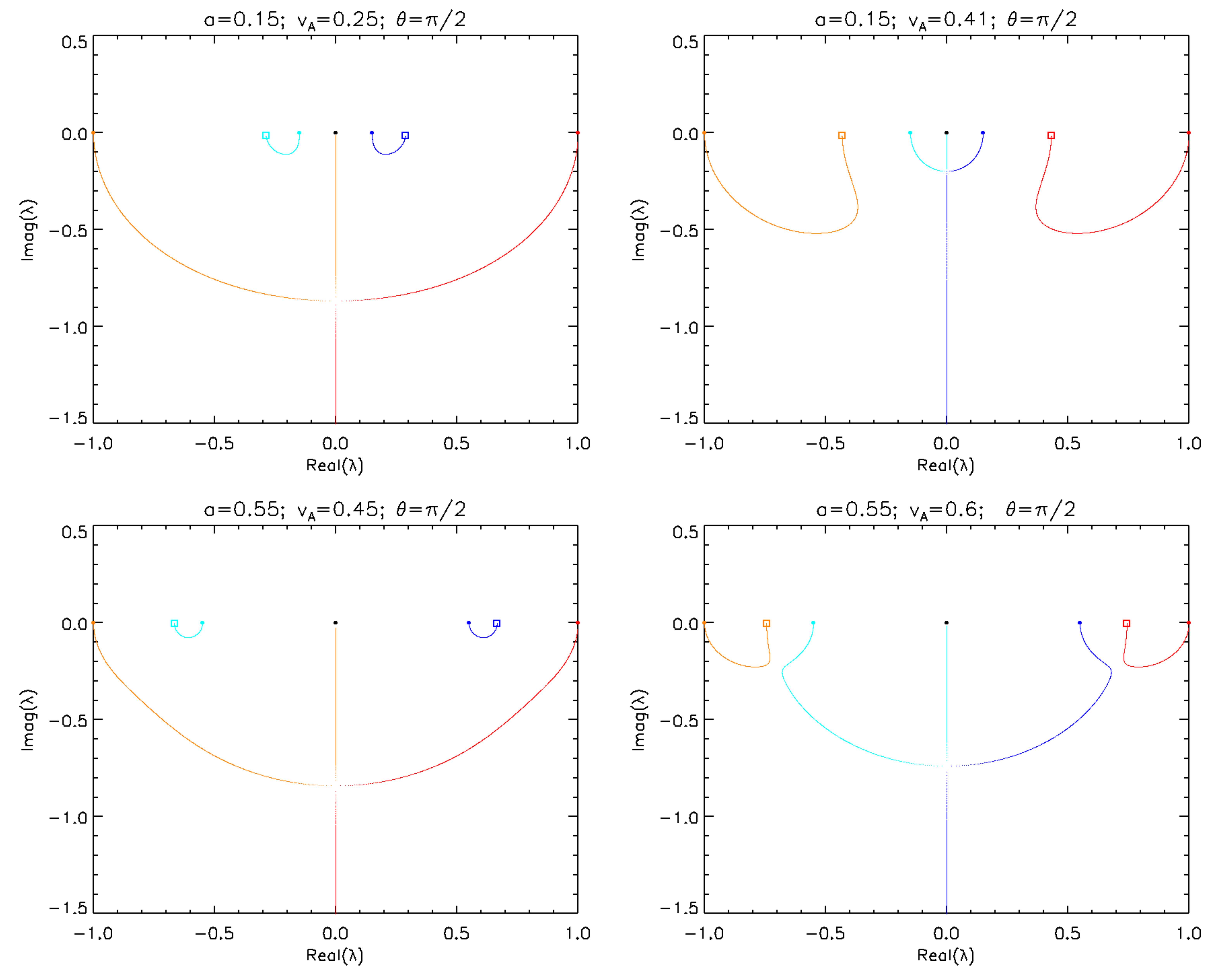}
  \caption{Roots of ${\cal P}_5$ in the complex plane for the
           perpendicular case ($\theta = \pi/2$):
           top panels correspond to the cold gas with $v_A = 0.25, 0.41$ while
           bottom panels refer to a hot gas and $v_A=0.55,0.6$
           At small (large) magnetizations - left (right) panels - acoustic
           (light) modes are non-degenerate and tend to the magnetoacoustic
           solution.
           Plotting conventions are the same one used throughout this paper.}
  \label{img:theta90}
\end{figure*}
Mode diagrams for different magnetizations $v_A=0.25,\, 0.41$ (cold gas) and $v_A= 0.45,\,v_A=0.6$ (hot gas) are illustrated in Fig. \ref{img:theta90}.
At small magnetizations (left panels), acoustic modes smoothly connect to the magnetoacoustic solution (\ref{eq:magnetoacoustic}) while light waves transition to a pair of \typeII modes.
At large magnetizations (right panels), light and acoustic modes reverse their asymptotic behaviors: the light degeneracy point disappears being replaced by the acoustic degeneracy point through which acoustic waves transition to a pair of \typeII modes.
By increasing $\tsigma$, one of these modes coincides with the rapidly damped mode (blue) while the second one (cyan) vanishes in the ideal limit.

The other four modes are given by the roots of
\begin{equation}
  \begin{split}
  {\cal P}^{\perp}_4 =& \lambda\Big[\lambda^3 + i\tsigma(u_A^2 + 2)\lambda^2 
   \\ \noalign{\medskip}
                     & -\left[(u_A^2+1)\tsigma^2 + 1\right]\lambda - 
                     i\sigma(u_A^2 + 1)\Big]
  \end{split}
\end{equation}
which have the simple analytical expressions:
\begin{equation}\label{eq:P4roots_perp}
  \lambda = \left\{\begin{array}{l}
    0
    \\ \noalign{\medskip}
    \DS \frac{1}{2}\left(-i\tsigma\pm\sqrt{4-\tsigma^2}\right)
    \\ \noalign{\medskip}
    -i\tsigma(u_A^2 + 1) \,.
  \end{array}\right.
\end{equation}
In this case there is a single light degeneracy (a root of multiplicity 2) always at $\tsigma = 2$ and it is independent of the magnetization.
A triple root is not physically admissible in this case.
The purely damped mode grows proportionally to $u_A^2$.

\subsubsection{Propagation at arbitrary angle $\theta$}
\begin{figure*}[!t]
  \centering
  \includegraphics[width=0.9\textwidth]{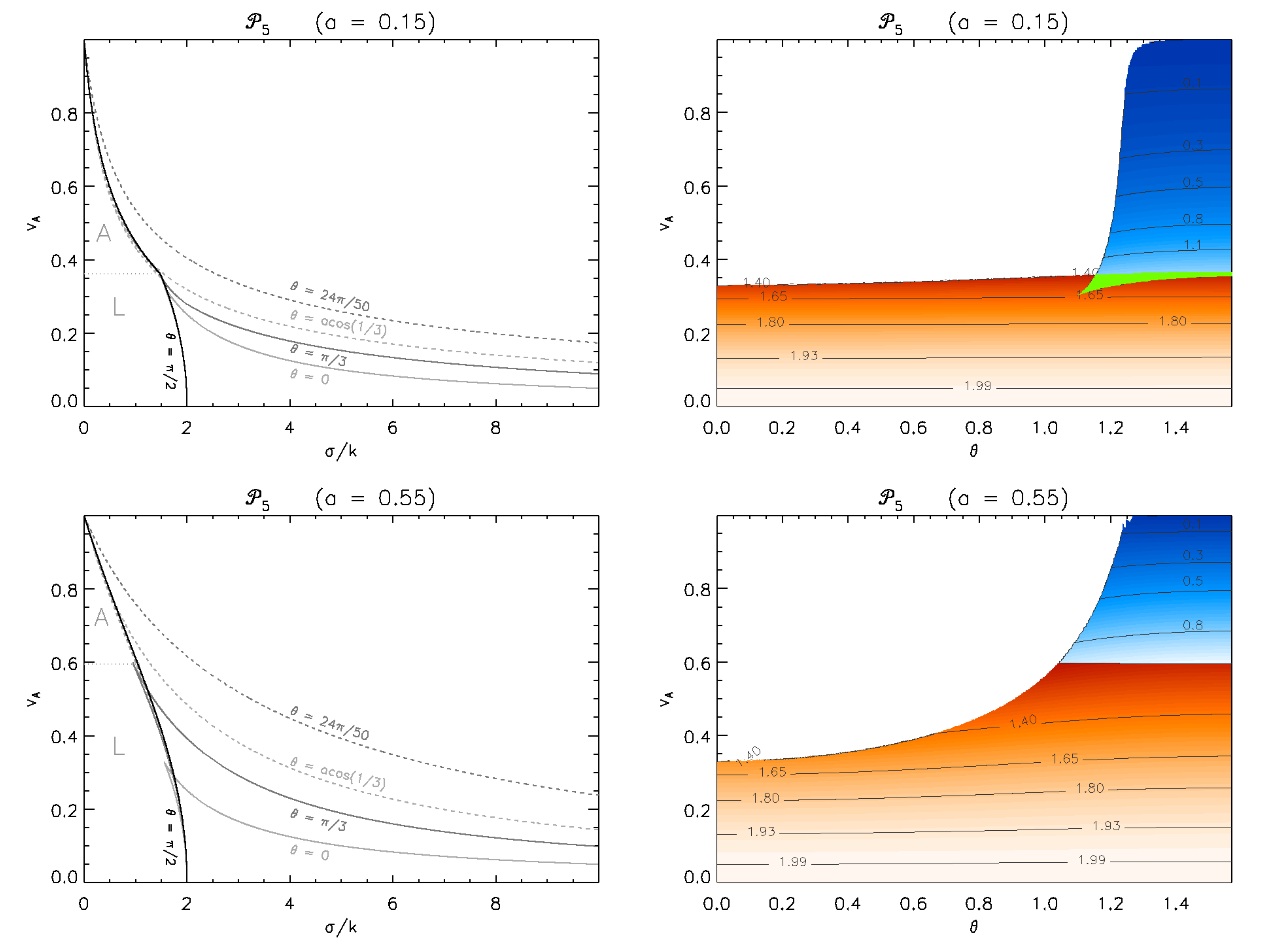}
  \caption{Eigenmode degeneracies of \Pfive for arbitrary angle in the cold
   ($a=0.15$, top panels) and hot ($a=0.55$, bottom panels) gas cases.
   The curves in the left panels show the values of $(\tsigma,\, v_A)$
   corresponding to a a root of multiplicity 2 and mark a transition from a \typeI to \typeII mode (left branch) or vice-versa (right branch).
   The cusp corresponds to the formation of a triple root.
   Light and dark-gray solid lines correspond to $\theta = 0$ and $\theta=\pi/3$.
   Similar dashed lines are used for $\theta=\cos^{-1}(1/3),\, 24\pi/50$. 
   In the right panel we show contour levels, in the $(\theta,v_A)$ plane,
   of $\tsigma$ at which the first degeneracy (\typeI-\typeII) occurs.
   Orange-filled levels correspond to the values of $\tsigma$ for which a light
   degeneracy point occurs, while blue-filled levels correspond to the acoustic
   degeneracy point.}
  \label{img:triplepoint_P5}
\end{figure*}

\begin{figure*}[!tp]
  \centering
  \includegraphics[width=0.9\textwidth]{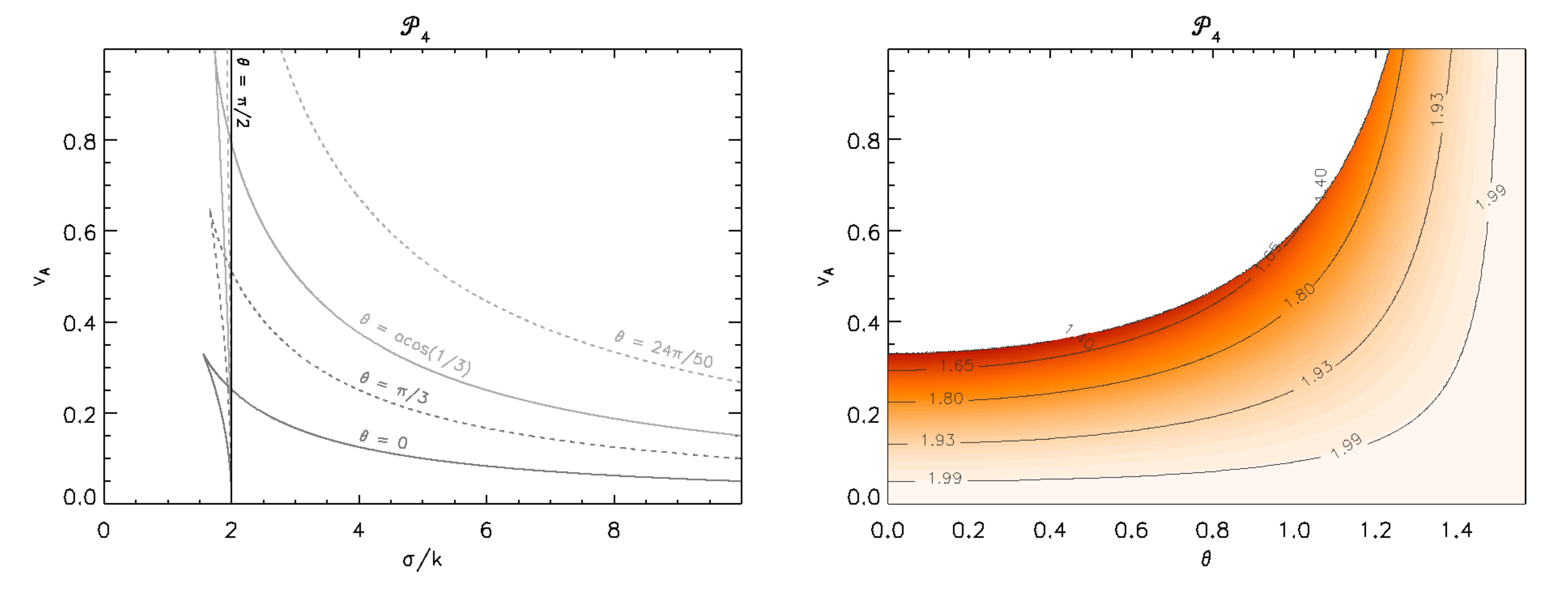}
  \caption{Eigenmode degeneracies of \Pfour for arbitrary angle propagation.
  The same plotting convention of Fig. \ref{img:triplepoint_P5} is used.}
  \label{img:triplepoint_P4}
\end{figure*}

Taking advantage of the results obtained in the previous sub-sections, we now explore the behavior at intermediate values of $\theta$.
The left panels in Fig. \ref{img:triplepoint_P5} show the locations of the degenerate roots for \Pfive for cold and hot gases (top and bottom plots on the left, respectively) in the $(\tsigma,\,v_A)$ plane for different values of $\theta$ (corresponding to different colored curves).

Inside each curve, a pair of \typeII mode exist; outside of this region, all roots (except the purely damped mode) are \typeI modes.
Across the leftmost branch of the curve, a root of multiplicity 2 sets the transition from \typeI to \typeII, typically a light or acoustic mode degeneracy.
Across the rightmost branch one has a transition from  \typeII to \typeI (e.g. slow/fast magnetosonic degeneracy).
Left and right branches intersect at a cusp point which marks the appearance of a \emph{triple} root (see Appendix \ref{app:P5perp_triple} for $\theta=\pi/2$).

The horizontal gray dotted line corresponds to the presence of a pair of double roots in the perpendicular case found at $u_A=\sqrt{a}$ (Eq. \ref{eq:P5perp_double_condition}).
As it will be shown shortly, this condition is nearly independent of $\theta$ and it will be used to separate the low magnetization region (where light waves may become degenerate, $u_A \lesssim \sqrt{a}$) from the high magnetization region (where acoustic waves may become degenerate, $u_A \gtrsim\sqrt{a}$).

The right panels in Fig. \ref{img:triplepoint_P5} employs color-filled contour levels to show the corresponding values of $\tsigma$, in the $\theta-v_A$ plane, at which the first degeneracy point is found.
Orange-filled contour levels correspond to light degeneracy points, i.e. transition from \typeI to \typeII modes.
Likewise, blue-filled levels indicate acoustic degeneracy points.
In the white region no degeneracy is present (all roots are distinct).
If a given value of $v_A$ and $\theta$ lies on a color-filled contour, then there exists a critical value of $\tsigma$ for which a degeneracy occurs.
This value is labeled by the corresponding contour level.
A \emph{triple} root exists at the boundary between a contoured and the white regions: cusp points on the left panel lie on this delimiting curve.

For $\theta=0$, degenerate roots are found only when $v_A <1/3$ (in correspondence of the two values of $\tsigma$ given by Eq. \ref{eq:theta0_double}).
This degeneracy affects only light modes (orange contours in the right panels), it does not depend on the sound speed and it is the same for \Pfive and \Pfour.
By increasing $\theta$ to $\pi/3$, the corresponding curve encloses a larger fraction of the parameter space the extent of which now depends on the value of the sound speed.
The cusp forms at larger values of $v_A$ ($v_A\approx 0.6$ in the hot gas case), as it is also clear from the right panels.
Results change significantly at larger angles ($\theta \gtrsim 1$): depending on the magnetization ($u_A \lesssim \sqrt{a}$ or $u_A \gtrsim\sqrt{a}$) either light or acoustic modes become degenerate for some value of $\tsigma$ as shown by the orange and blue contours in the right panels, respectively.
An overlapping region where both light and acoustic waves become \typeII modes exists for the cold gas case (green area in the top right panel).
As $\theta$ approaches $\pi/2$ (perpendicular propagation), a degeneracy takes place at any magnetization (dashed curves in the left panels in Fig. \ref{img:triplepoint_P5}).
In the limiting case $\theta=\pi/2$ the rightmost branch of the curve becomes horizontal and stretches out to $\tsigma=\infty$ indicating the disappearance of the slow modes.

The previous discussion can be extended to the roots of ${\cal P}_4$ using the same plotting conventions.
From the left and right panels in Fig. \ref{img:triplepoint_P4}, it is seen that light modes always suffer from a degeneracy (\typeI - \typeII transition) at some critical value of $\tsigma$ in the two following cases:
\begin{itemize}

\item
For any $\theta\in[0,\pi/2]$ and $v_A < 1/3$ (weak magnetizations).
This is a weak condition since the value of $v_A=1/3$ provides only a lower bound which we know from the case of parallel propagation (see the discussion after Eq. \ref{eq:theta0_double}).
The region extends indeed to larger values of $v_A$ as $\theta$ is increased.
      
\item
For $\theta \ge \theta_1 \equiv \cos^{-1}(1/3)$ and any value of $v_A$.
The value $\theta_1$ corresponds to the intersection point between the orange-white demarcation line and the $v_A=1$ axis.
The exact value of $\theta_1$ can be found by writing ${\cal P}_4$ in the limit of strong magnetization ($v_A=1$),
\begin{equation}
  \lim_{u_A\to\infty} {\cal P}_4 =   i\tsigma\lambda^3 - \tsigma^2\lambda^2
                                   - i\tsigma\lambda   + \tsigma^2\cos^2\theta \,,
\end{equation}
and by imposing the condition for a perfect cubic (triple root).
This yields $\cos\theta_1 = 1/3$ and $\tsigma_1=\sqrt{3}$ and corresponds to the cusp point brushing the $v_A=1$ axis in the left panel of Fig. \ref{img:triplepoint_P4} (light grey solid line).
Thus, for strongly magnetized plasmas ($v_A\sim 1$) light modes propagating almost perpendicularly become degenerate for some value $\tsigma \ge \sqrt{3}$.
\end{itemize}
The second degeneracy, corresponding to the \typeII-\typeI transition (rightmost branch in the left panels in Fig. \ref{img:triplepoint_P4}) shifts at increasingly larger values of $\tsigma$ and it extends to infinity as $\theta\to\pi/2$.

\subsubsection{Polar Diagram.}

\begin{figure*}[!pt]
  \centering
  \includegraphics[width=0.99\textwidth]{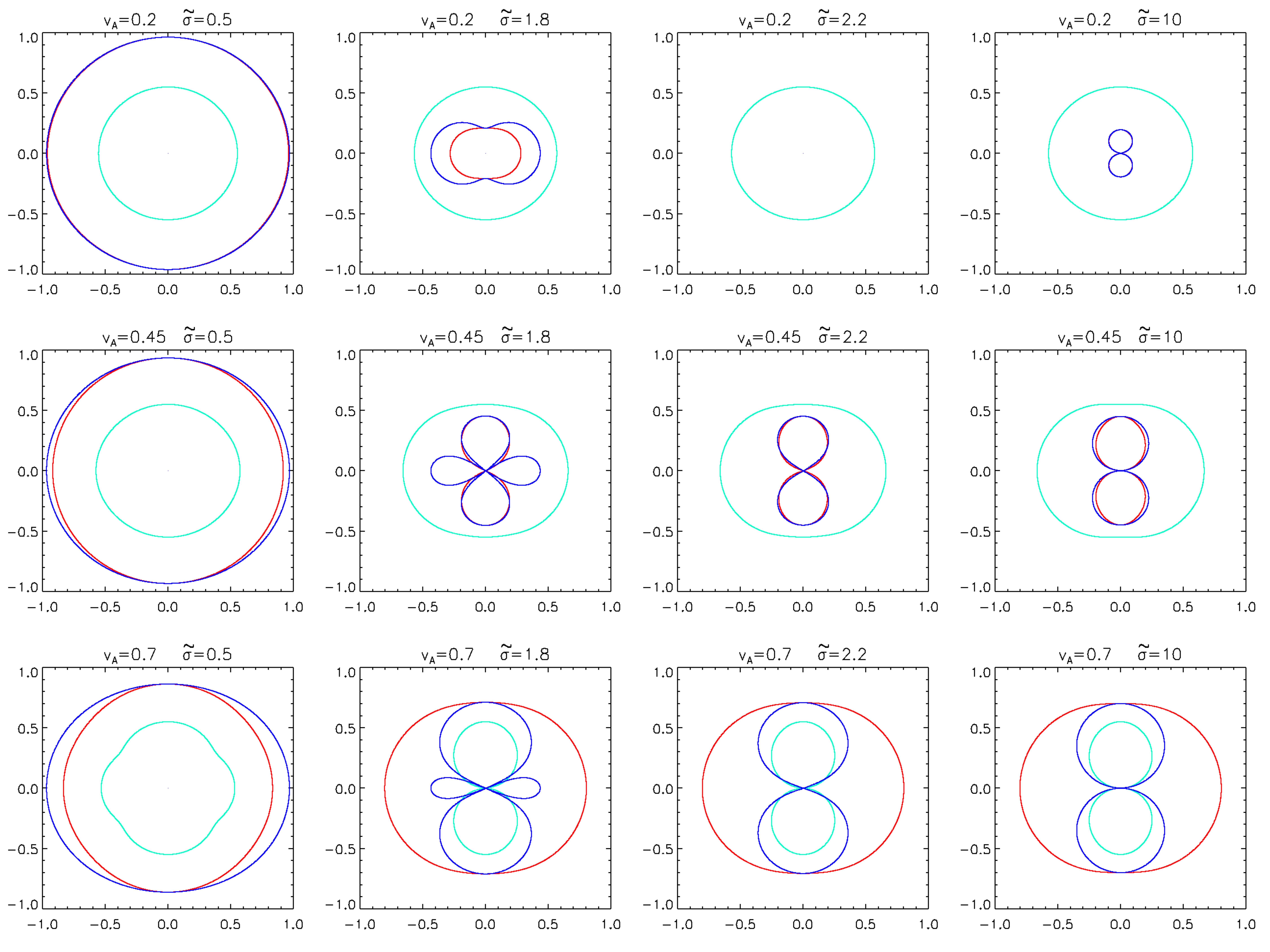}
  \caption{\footnotesize Polar diagrams showing the phase velocity for different
  value of $\tsigma$ (left to right) and of the magnetization parameter $v_A$
  (top to bottom).}
  \label{img:polar}
\end{figure*}

The phase velocity of the waves can be plotted as a function of the polar angle measured from the direction of the background field $B_0$.
Since our results are only weakly depending on the value of the sound speed, we now restrict our attention to $a=0.55$.
The most prominent cases are shown in the sequence of panels Fig. \ref{img:polar} where polar diagrams for the roots of \Pfive and \Pfour are shown using green, red (for the former) and blue (for the latter).
From left to right, we show a sequence of panels corresponding to increasing values of $\tsigma$.
From the previous discussion, a \typeI-\typeII transition is expected around $\tsigma\approx 2$ for a weakly magnetized plasma.
For this reason, selected plots are shown using values of $\tsigma$ immediately before and after this transition threshold.

\begin{itemize}
\item
For small values of the conductivity ($\tsigma=0.5$, leftmost panels in Fig. \ref{img:polar}) signal velocities of light and acoustic modes propagate essentially isotropically with a weak dependence on the angle.
The light-waves of \Pfour are slightly larger than those of \Pfive but they coincide in the case of parallel propagation ($\theta=0$), as also shown by Eqs. (\ref{eq:P4_theta0}) and (\ref{eq:P5_theta0}).

\item
At $\tsigma=1.8$ (second column of panels), no degeneracy is yet present for $v_A=0.2$ and the phase speed of the light modes becomes smaller than the sound speed.
When the magnetization is increased at $v_A=0.45$, light-waves of \Pfive become degenerate in a narrow range around $\theta \approx \pi/3$ (see the bottom left panel in Fig. \ref{img:triplepoint_P5}) whereas acoustic waves propagate distinctly.
Finally, when $v_A = 0.7>\sqrt{a/(1+a)}$, light modes are distinct and the acoustic mode are now degenerate.

In the case of \Pfour, light modes become first degenerate at some intermediate value of $\theta$ ($1\lesssim \theta\lesssim 1.2$, see the right panel in Fig. \ref{img:triplepoint_P4}) while roots are again distinct for larger values of $\theta$.

\item
For $\tsigma=2.2$ (third column of panels), only the acoustic modes can propagate at small magnetization ($v_A=0.2$, top) while all light modes have become \typeII modes.
Increasing the magnetization to $v_A=0.45$ (second panel from top), we see that light modes can propagate parallel to the field but become suppressed in a narrow range around $\theta \gtrsim \pi/3$.
Strengthening the field to $v_A=0.7$ leads to the degeneracy of the acoustic modes and the \Pfour light modes at large angles while light-waves of \Pfive are, as expected, distinct.

\item
For $\tsigma=10$ (rightmost column of panels), we recover the usual ideal polar diagram for fast, slow and Alfv\'en waves.
Fast and slow magnetosonic modes are given by the roots of \Pfive while Alfv\'en waves are given by the roots of \Pfour.
For weak and moderate magnetizations (first and second panels from the top) the green curve identifies the fast mode (this solution is always smoothly connected to the acoustic mode) while red and blue curves are very similar and represent pairs of slow and Alfv\'en modes (no perpendicular propagation is allowed for these solutions).
This trend reverses once the magnetization is strong enough ($v_A=0.7$, third panel from the top) because of the light-acoustic degeneracy: the light modes of \Pfive (red) have now become fast magnetosonic waves whereas blue and green identity, respectively, pairs of Alfv\'en and slow magnetosonic modes.
\end{itemize}

\subsection{Group, Signal and Front Velocity}
\label{sec:signal_velocity}
%
%
%

The results of the previous sections raise some interesting questions about the significance of the group velocity.
Being the medium dissipative, the classical expression for the group velocity $v_g=d\omega/d k$ (see Eq. \ref{eq:group_vel}) is complex, so a first question is about its physical meaning.
This issue has been adressed by \cite{Muschietti_Dum_1993} who showed that, because the wavenumber components are damped at different rates, the central wavenumber changes with time. 
The imaginary part of the group velocity accounts for this change.

\begin{figure}[!pt]
  \centering
  \includegraphics[width=0.45\textwidth]{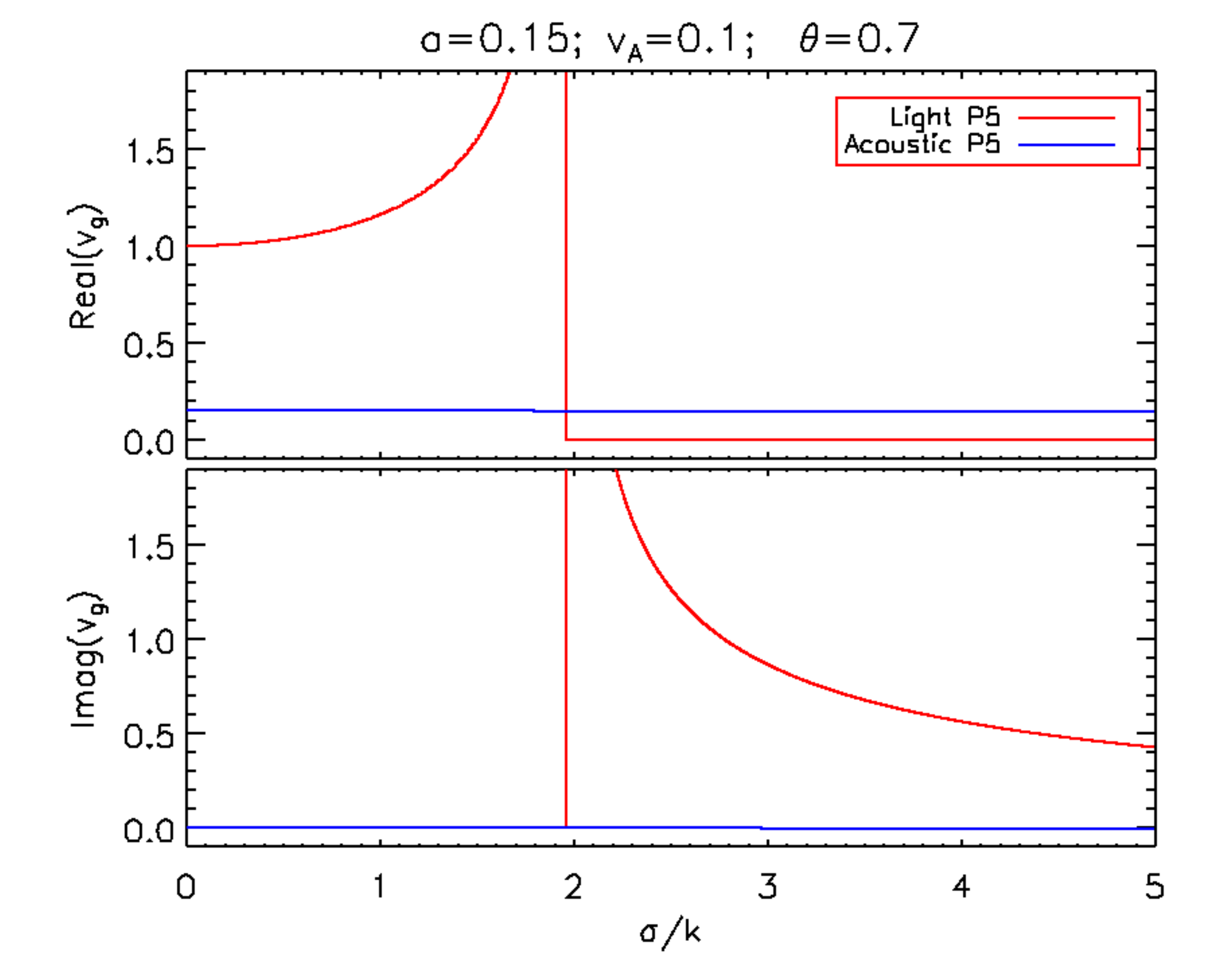}
  \caption{\footnotesize Group velocities for \Pfive corresponding
  to the same parameters used in Fig. \ref{img:cold_lowM}.
  Only the upper-half plane in the region $\tsigma\in[0,5]$ is shown.}
  \label{img:group_P5}
\end{figure}

A second question arises because (the real part of) $v_g$ may occasionally exceed unity when the real part of $\lambda$ quickly approach zero at degenerate points or for small conductivities.
An example, using $\theta=0.7$, $a = 0.1$ and $v_A=0.1$ is shown in Fig. \ref{img:group_P5} where we plot the group velocity for the light and acoustic waves.
We remind, however, that the group velocity represents the propagation speed of an envelope which is not too broad in wave number but, in general and contrary to a diffuse misconception, it  \emph{does not} define the speed at which information travels (see, for instance, \cite{Stratton_1941} page 337,  \cite{Brillouin_1960}, \cite{Jackson_1998} page 324).
The actual signal velocity, instead, is related to the propagation of a wave packet with finite spatial width \cite{Sommerfeld_1960} or to a short isolated succession of wavelets, with the system being at rest before the signal arrives and also after it has passed \cite{Brillouin_1960}.
In this respect, a closely related concept is that of the \emph{front velocity} which tracks the very first arrival of a disturbance that carries information that cannot be predicted from an earlier time.
Causality cannot be violated if the front velocity is less or equal to the speed of light.

\begin{figure}[!pt]
  \centering
  \includegraphics[width=0.45\textwidth]{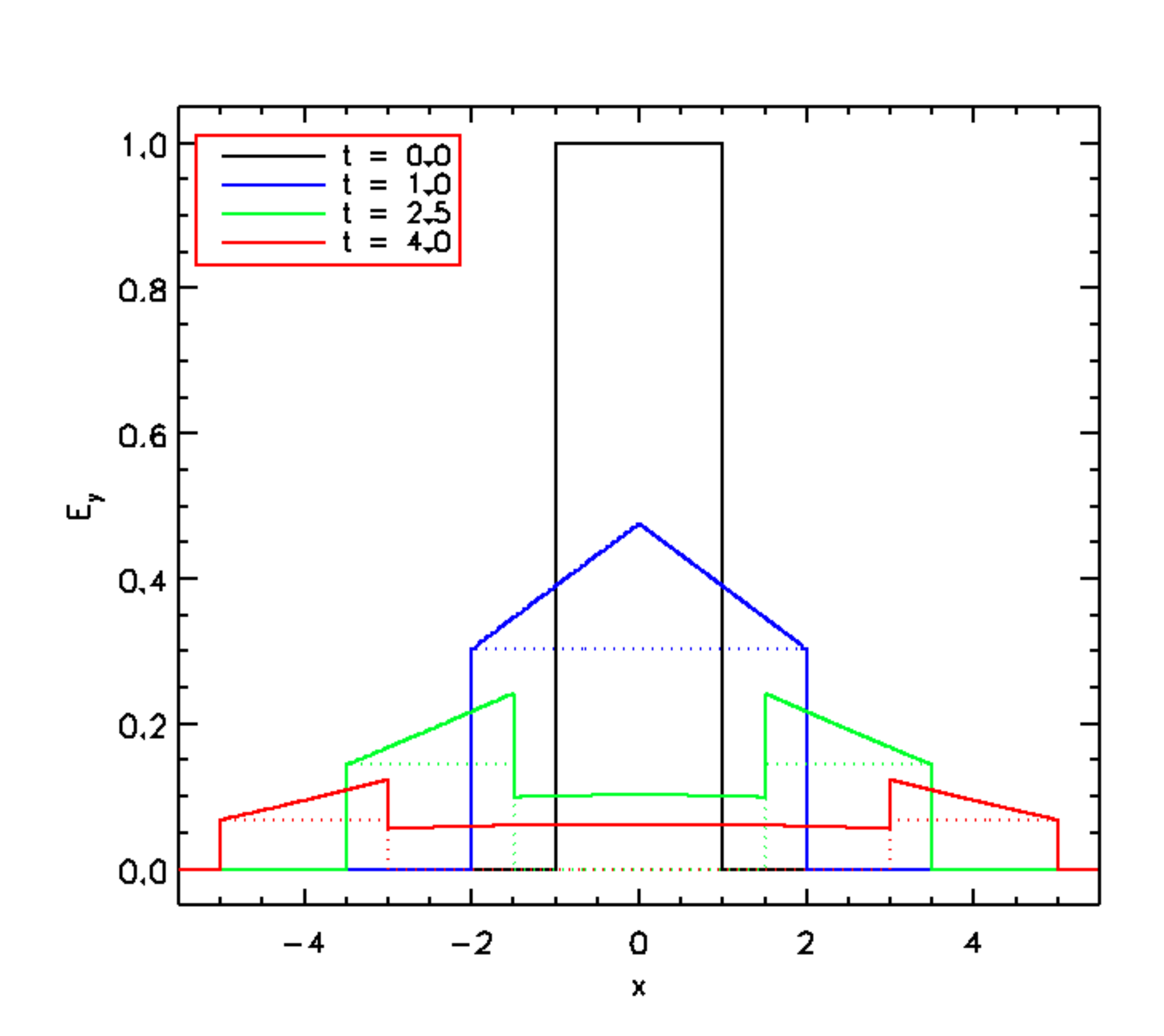}
  \caption{\footnotesize Evolution of a square pulse in a dissipative
  dispersive media with dispersion relation given by
  $\omega(k) = -\frac{i}{2\eta} \pm \sqrt{k^2 - \frac{1}{4\eta^2}}$
  (see Eq. \ref{eq:P4roots_perp}) where $\eta = 1/\sigma = 2$.
  The dotted line give the corresponding solution in an ideal medium
  ($\omega = k$).}
  \label{img:signal_velocity}
\end{figure}
To this purpose we consider the special case of perpendicular propagation (section \ref{sec:perp_prog}) for which the dispersion relation has analytical expressions given by Eq. (\ref{eq:P4roots_perp}).
From that expression, the group velocity is found to be
\begin{equation}
 v_g(k) = \pm\frac{2}{\sqrt{4-\tsigma^2}} = \pm\frac{2k}{\sqrt{4k^2 - \sigma^2}} \,,
\end{equation}
which is always superluminal and even diverges for $\tsigma \to 1/2$.
Note also, that the previous expression coincides with the expression given in Section IV of \cite{Koide_2008}.
It is easy to show that the equations for the $E_y$ and $B_z$ reduce to the telegraph equation,
\begin{equation}\label{eq:telegraph}
  \pd{^2\psi}{t^2} + \sigma \pd{\psi}{t} - \pd{^2\psi}{x^2} = 0 \,.
\end{equation}
where $\psi\equiv\psi(x,t)$ stands for either $E_y$ or $B_z$.
An harmonic analysis in space (see section [5.10] of \cite{Stratton_1941}) shows that the solution of Eq. (\ref{eq:telegraph}) is determined by the wavefunction
\begin{equation}\label{eq:wave_function}
  \begin{split}
  \psi(x,t) = \frac{e^{-\sigma t/2}}{2}\Big[ & \psi_0(x+t) + \psi_0(x-t)
     \\
          &   + \frac{\sigma}{2}{\cal D}_0(x,t) + {\cal D}_1(x,t) \Big]
  \end{split}
\end{equation}
where $\psi_0(x) = \psi(x,0)$ is the initial condition while the ${\cal D}_n$ terms are integrals of the Bessel function of the first kind and its derivative times the initial distribution, 
\begin{equation}
  \begin{array}{lcl}
   {\cal D}_0(x,t) &=&\DS \int_{x-t}^{x+t} \psi_0(\xi)
                       J_0\big(z(x,t,\xi)\big) d\,\xi
    \\ \noalign{\medskip}
   {\cal D}_1(x,t) &=&\DS \int_{x-t}^{x+t} \psi_0(\xi)
                  \pd{}{t}J_0\big(z(x,t,\xi)\big) d\,\xi
  \end{array}
\end{equation}
with $z(x,t,\xi)=(\sigma/2) \sqrt{(x-\xi)^2 - t^2}$
(in our derivation we have set the term $\left.\partial\psi/\partial t\right|_{t=0} = 0$).
For an initial square pulse $\psi_0(x) = \big(1+{\rm sgn}(x_0-|x|)\big)/2$, the wavefunction given by Eq. (\ref{eq:wave_function}) has been computed numerically and it is plotted in Fig. \ref{img:signal_velocity} at different times $=0,1,2.5,4$.
For this calculation $\sigma = 1/2$ has been used.
The evolution discloses that the initial distribution splits into a pair of damped, left- and right-going waves.
The contribution of the integrals ${\cal D}_n$ does not alter the propagation speed (the integral vanishes for $|x|>|x_0| + t$) but it deformates the shape of the wave leaving a residue field after the front has passed through.
The speeds of the two fronts always remain equal to the speed of light ($=1$).

\section{Summary}
\label{sec:summary}
%
%
%

A characteristic analysis of the resistive relativistic MHD equations has been the subject of this work.
Starting from an equilibrium state describing a static and homogeneous relativistic plasma threaded by a constant magnetic field, perturbations have been introduced in the form of plane waves $\propto\exp[ i(kx-\omega t)]$, where $k\in\mathbb{R}$ while $\omega\in\mathbb{C}$ is a complex quantity.
The dispersion relation has been obtain as a ten-degree polynomial which can be factorized into a single root $\lambda=0$ and two lower-order polynomials of degree five and four, respectively.
The coefficients of the two polynomials are expressed in terms of four parameters: the sound speed $a$, the magnetization $u_A = B_0/\sqrt{w_0}$ (or $v_A=u_A/\sqrt{1+u_A^2}$), the angle $\theta$ between the wavevector and the background magnetic field $B_0$ and the ratio $\sigma/k$ between the electric conductivity $\sigma$ and the wavenumber $k$.

Solution modes are of two kinds: i) waves with non-zero phase speed which always come as pairs of opposite complex conjugate solutions or ii) purely damped standing waves.
The isolated root $\lambda=0$ coincides with the contact mode and it is unaffected by resistivity.
The remaining waves can be easily identified in the fully resistive limit (zero conductivity or small wavelengths) where electromagnetic fields and matter are decoupled so that characteristic information is propagated through light or sound waves.
In this limit one has four light-waves, two acoustic waves and three damped waves (in addition to the contact mode).
In the ideal limit (infinite conductivity or large wavelengths), solution modes asymptotically approach pairs of fast, slow or Alfv\'en waves (and the contact mode).
Using asymptotic analysis we have shown that the damping rates of these propagating modes scales as $\eta k^2$ ($\eta$ is the plasma resistivity), as expected for a diffusive system.
Conversely, the three damped modes become singular solutions of the equations and become linearly suppressed with the conductivity.

For arbitrary values of $\sigma/k$, the dispersion relation cannot be solved in closed analytical form and a numerical approach has been employed.
Our results confirm that eigenvalues are, in general, complex quantities with negative (or zero) imaginary part indicating wave damping, a defining feature of dissipative systems.
Given the nonlinear dependency on $\sigma/k$, the system is also dispersive with light waves propagating at small wavelengths while fast or slow mode propagating at large wavelengths.

In general, the solution space is characterized by a number of mode transitions which involve a root degeneracy.
Isolated roots of multiplicity two define a boundary region of the parameter space inside which a pair of propagating (\typeI) modes has transitioned to a pair of non-propagating (\typeII) modes.
Conversely, through a pair of double roots, solution modes switch their asymptotic behavior (e.g., light and acoustic waves interchange with each other) by remaining \typeI modes.
These transition points are described by degeneracy conditions of quintic and quartic polynomials and, in general, no simple expression have been found except for special cases.
However, some general results could be established:

\begin{itemize}

\item
For weak magnetization - namely $u_A < 1/\sqrt{8}$ for parallel propagation or $u_A \lesssim \sqrt{a}$ at larger angles - there is always a finite range of values of $\sigma/k$ where light modes degenerate into a pair of standing damped waves.
On the contrary, acoustic modes remain distinct for any value of $\sigma/k$ and, in the ideal limit they asymptotically approach the fast (when $v_A \lesssim a$) or slow magnetosonic (when $v_A \gtrsim a$) waves.

\item
For sufficiently stronger magnetizations and $\cos\theta \lesssim 1/3$, no degeneracy occurs and the four light-waves and the two acoustic modes smoothly connect to fast, slow and Alfv\'en waves in the ideal limit.
The magnetization threshold coincides with $u_A=1/\sqrt{8}$ for parallel propagation but it increases with the inclination angle.

\item 
As the inclination becomes more perpendicular ($\cos^{-1}(1/3)\lesssim\theta\le\pi/2$) and $u_A \gtrsim \sqrt{a}$, only two light-waves remain distinct while the remaining \typeI solutions (2 acoustic and 2 light modes) always become degenerate for some intermediate value range of $\sigma/k$.
In the limit of very strong magnetic fields, acoustic modes become quickly suppressed and disappear for perpendicular propagation.
In the limit $\sigma/k\to\infty$ the two distinct roots smoothly connect to the fast magnetosonic modes while the remaining ones tend to slow and Alfv\'en solutions.
\end{itemize}

To the extent of our knowledge, our results provide the first systematic analysis of the characteristic structure of the relativistic MHD equations in presence of a finite conductivity.
The outcome of this work may be particularly relevant in the field of relativistic magnetic reconnection as well as representing a potential benefit for the development of improved numerical methods in the solution of these kind of equations.

\appendix

\section{Purely Imaginary Solutions of the Dispersion Relation}
\label{app:imag_sol}
%
%
%

Here we show that \Pfive{ } always admits at least one \typeII (purely imaginary) solutions while, in the case of \Pfour{ }, at least two solutions of this type must be present.

\paragraph{Proof for ${\cal P}_5$}.
We seek for a solution of the type $\lambda = iY$ in Eq. (\ref{eq:P5}).
Hence it is readily found, from Eq. (\ref{eq:P5}) that 
\begin{equation}\label{eq:P5_imag}
  \begin{split}
  {\cal P}_5(iY) =&\quad i\Big[Y^5
              + \tsigma(u_A^{2} + 1)Y^4
              + (a^2 + 1)Y^3                 \\
              & \tsigma(C^2 + a^{2} + u_A^{2})Y^2 + a^{2}Y + \tsigma C^2\Big]
  \end{split}              
\end{equation}
where $C = a^{2}u_A^{2}\cos^{2}\theta$.
The polynomial inside the square brackets is a real-valued quintic function which must always possess at least one real root.
Thus $\lambda = iY$ is a purely imaginary solution of the original polymomial.

\paragraph{Proof for ${\cal P}_4$}.
Similarly, we seek for a solution of the type $\lambda = iY$ in Eq. (\ref{eq:P4}).
Upon substituting in Eq. (\ref{eq:P4}) we find
\begin{equation}\label{eq:P4_imag}
  \begin{split}
  {\cal P}_4(iY) =&\quad  Y^4 + \tsigma(u_A^2 + 2)Y^3
                      + \Big[(u_A^{2} + 1)\sigma^2 + 1\Big]Y^2  \\
                  &   + \tsigma(u_A^2 + 1)Y
                      + \tsigma^{2}u_A^2\cos^{2}\theta \,.
  \end{split}                          
\end{equation}
The previous equation is again a real-valued quartic equation which has the following properties:
\begin{equation}
  \left\{\begin{array}{lcl}
   \DS \lim_{Y\to-\infty} {\cal P}_4(iY) &=& +\infty  \\ \noalign{\medskip}
   \DS {\cal P}_4(0)        &=& \tsigma^2u_A^2\cos^2\theta  \\ \noalign{\medskip}
   \DS {\cal P}_4(-i\tsigma) &=& -\tsigma^2u_A^2\sin^2\theta  
  \end{array}\right.
\end{equation}
For $\theta>0$ the quartic is positive at $Y\to-\infty$ and $Y=0$ but negative in the neighbourhood of $Y=-\tsigma$ and thus (at leat) two roots must be found in the range $Y\in[-\infty,0]$ which proves our statement.
In the special case $\theta = 0$, the quartic simplifies to
\begin{equation}
  (Y+\tsigma)\left[Y^3 + \tsigma(u_A^2+1)Y^2 + Y + \tsigma u_A^2\right] = 0
\end{equation}
which is satisfied for $Y=-\tsigma$ and by at least one root of the cubic inside the square brackets.

\section{Eigenvectors Expression in the Resistive and Ideal Limits}
\label{app:eigenvectors_limits}
%
%
%

In the $\tsigma\to 0$ (resistive) limit, Eq. (\ref{eq:eigenvectors_comp}) can still be used to obtain the eigenvectors for the compressible modes which, not surprisingly, reduce to a pair of relativistic sound waves carrying perturbations in density, pressure and normal velocity only:
\begin{equation}
\left(\begin{array}{l}
  \rho_1       \\ \noalign{\medskip}
  v_{1x}       \\ \noalign{\medskip}
  p_1
  \end{array}\right)
  =
  \left(\begin{array}{c}
  1
  \\ \noalign{\medskip}
  \frac{a}{\rho_0}
  \\ \noalign{\medskip}
  \frac{a^2w_0}{\rho_0}
  \end{array}\right) \,.
\end{equation}
However, for the light modes, the assumption $\rho_1= 0$ leads to a singular expression but the direct solution of Eq. (\ref{eq:linear_system}) with $\rho_1=0$ provides the usual eigenvectors for Maxwell equations:
\begin{equation}
\left(\begin{array}{l}
  B_{1y}       \\ \noalign{\medskip}
  B_{1z}       \\ \noalign{\medskip}
  E_{1x}       \\ \noalign{\medskip}
  E_{1y}       \\ \noalign{\medskip}
  E_{1z}       \\ \noalign{\medskip}
  \end{array}\right)
   =
\left(\begin{array}{c}
  1
  \\ \noalign{\medskip}
  0
  \\ \noalign{\medskip}
  0
   \\ \noalign{\medskip}
  0
   \\ \noalign{\medskip}
  \pm 1
  \end{array}\right) \,,
\left(\begin{array}{c}
  0
  \\ \noalign{\medskip}
  1
  \\ \noalign{\medskip}
  0
   \\ \noalign{\medskip}
  \pm 1
   \\ \noalign{\medskip}
  0
  \end{array}\right) \,,
\end{equation}
where $\rho_1=\vec{v}_1 = p_1 = 0$.

In the ideal limit (large wavelenghts or infinite conductivity), ideal limit), the compressible modes are given by Eq. (\ref{eq:eigenvectors_comp}) by simply taking $\sigma/k \to \infty$:
\begin{equation}\label{eq:eigenvectors_comp_ideal}
\left(\begin{array}{l}
  \rho_1       \\ \noalign{\medskip}
  v_{1x}       \\ \noalign{\medskip}
  v_{1y}       \\ \noalign{\medskip}
  v_{1z}       \\ \noalign{\medskip}
  B_{1y}       \\ \noalign{\medskip}
  B_{1z}       \\ \noalign{\medskip}
  E_{1x}       \\ \noalign{\medskip}
  E_{1y}       \\ \noalign{\medskip}
  E_{1z}       \\ \noalign{\medskip}
  p_1
  \end{array}\right)
   =  \rho_1
\left(\begin{array}{c}
  1
  \\ \noalign{\medskip}
  \frac{\lambda}{\rho_0}
  \\ \noalign{\medskip}
  -\frac{\lambda u_A^2\sin\theta\cos\theta(1-\lambda^2)}{\rho_0\Delta}
   \\ \noalign{\medskip}
   0
   \\ \noalign{\medskip}
  \frac{\lambda^2\sqrt{w_0}u_A\sin\theta}{\rho_0\Delta}
   \\ \noalign{\medskip}
   0
   \\ \noalign{\medskip}
   0
   \\ \noalign{\medskip}
   0
   \\ \noalign{\medskip}
   -\frac{ \lambda^3\sqrt{w_0}u_A\sin\theta }{ \rho_0\Delta }
   \\ \noalign{\medskip}
    \frac{ \lambda^2w_0((u_A^2+1)\lambda^2-u_A^2) }{ \rho_0\Delta } 
  \end{array}\right)
\end{equation}
where now $\Delta = \lambda^2 - (1-\lambda^2)u_A^2\cos^2\theta$ while $\lambda$ is given by the fast and slow modes (Eq. \ref{eq:fast_and_slow}).
Incompressible perturbations are instead given by
\begin{equation}\label{eq:eigenvectors_incomp_ideal}
\left(\begin{array}{l}
  \rho_1       \\ \noalign{\medskip}
  v_{1x}       \\ \noalign{\medskip}
  v_{1y}       \\ \noalign{\medskip}
  v_{1z}       \\ \noalign{\medskip}
  B_{1y}       \\ \noalign{\medskip}
  B_{1z}       \\ \noalign{\medskip}
  E_{1x}       \\ \noalign{\medskip}
  E_{1y}       \\ \noalign{\medskip}
  E_{1z}       \\ \noalign{\medskip}
  p_1
  \end{array}\right)
   =  B_{1z}
\left(\begin{array}{c}
  0
  \\ \noalign{\medskip}
  0
  \\ \noalign{\medskip}
  0
   \\ \noalign{\medskip}
  -\frac{\lambda\tsigma}{B_0\cos\theta}
   \\ \noalign{\medskip}
  0
   \\ \noalign{\medskip}
  1
   \\ \noalign{\medskip}
  -\lambda \tan\theta
   \\ \noalign{\medskip}
  \lambda
   \\ \noalign{\medskip}
   0
   \\ \noalign{\medskip}
   0
  \end{array}\right) \,.
\end{equation}
where now $\lambda$ is given by the Alfv\'en modes, Eq. (\ref{eq:Alfven}).
Moreover, Eq. (\ref{eq:eigenvectors_incomp_ideal}) reduces to the the classical MHD expressions in the non-relativistic limit $w_0\to\rho_0$ where $v_{1z} = \mp 1/\sqrt{\rho_0}$, $B_{1z} = 1$ and $\vec{E}_1 = -\vec{v}_1\times\vec{B}_0$.

\section{Triple Root of \Pfive in the Perpendicular Case}
\label{app:P5perp_triple}
%
%
%

We now discuss the degenerate roots of \Pfive in the perpendicular case.
From Eq. (\ref{eq:P5perp}), the quartic polynomial inside square bracket can be converted to depressed form using the substitution $\lambda = i(Y-c_3/4)$, where $c_3 = \tsigma(u_A^2+1)$.
This yields
\begin{equation}
  f(Y) = Y^4 + a_2Y^2 + a_1Y + a_0
\end{equation}
where
\begin{equation}
  \begin{array}{lcl}
    a_2 & = & \DS -\frac{3}{8}(u_A^2+1)^2 + a^2 + 1 \\ \noalign{\medskip}
    a_1 & = & \DS \frac{\tsigma(u_A+1)}{2}\left[
                \frac{1}{4}\tsigma^2(u_A+1)^2
              + (u_A-1)(1-a^2) \right]\\ \noalign{\medskip}
    a_0 & = & \DS -\frac{3}{256}(u_A^2+1)^4\tsigma^4
              +\frac{1}{16}(u_A^2+1)\left[1 - 3a^2 + \right.
             \\ \noalign{\medskip}
        &   & \left. (a^2-3)u_A^2\right]\tsigma^2 + a^2
  \end{array}
\end{equation}
Written in this form, the condition to have a triple root (see, for instance, \cite{Viher_2007}) is:
\begin{equation}
  a_2^2 + 12a_0 = 0 \quad{\rm and}\quad 8a_2^3 + 27a_1^2 = 0
\end{equation}
together with $a_2 < 0$.
The first of the two conditions can be readily solved for $\tsigma^2$ yielding
\begin{equation}\label{eq:P5perp_triple_sigma}
  \tsigma^2 = \frac{1}{3}\frac{a^4 + 14a^2 + 1}{(u_A^2 + 1)(u_A^2 + a^2)}
\end{equation}
and then inserted into the second one, giving the following biquadratic equation for $u_A$:
\begin{equation}\label{eq:P4red_uA}
  \begin{split}
     & +  8\Big[a^6 + 3a^2(a^2 + 1) + 1\Big] u_A^4 \\
     & -  \Big[a^8 + 76a^2(a^4+1) - 282a^4 + 1\Big]u_A^2  \\
     & + 8a^2\Big[a^6 + 3a^2(a^2 + 1) + 1\Big] = 0
  \end{split}                          
\end{equation}
Apart from the tedious form of the coefficients, the solution can be written 
\begin{equation}\label{eq:uA_triple}
  u_A^2 = \frac{1}{16} \frac{B \pm (1-a^2) C^{3/2}}{(a^2+1)^3}
\end{equation}
where $B=a^8 + 76a^2(a^4+1) - 282a^4 + 1$, $C = (a^2+1)^2 - 36a^2$
Physically admissible solutions require the argument of the square root to be positive, that is, $0\le a \le 3-\sqrt{8}$.
The location of the triple roots is shown by the red and orange curves in Fig. \ref{img:P5perp_triple}.

\bibliographystyle{aipnum4-1}
\bibliography{paper}

\end{document}